\documentclass[a4paper,10pt]{article}
\usepackage{amsmath,amsfonts,amssymb,amsthm,graphics,graphicx,epsfig,times,bbm,verbatim}

\usepackage[utf8]{inputenc}
\usepackage{authblk}

\newtheorem{definition}{Definition}

\newtheorem{lemma}{Lemma}

\newtheorem{example}{Example}

\newtheorem{theorem}{Theorem}

\usepackage{amsfonts}
\usepackage{psfrag}
\usepackage{dsfont}
\usepackage[dvips]{rotating}
\usepackage{epsfig}
\usepackage{pstricks,pst-grad}

\newcommand{\md}{\mathrm{d}}
\newcommand{\Tr}{\mathrm{Tr}}

\newcommand{\id}{\mathbbm{1}}
\newcommand{\EE}{\mathbbm{E}}
\newcommand{\PR}{\mathbbm{P}}
\newcommand{\PP}{\mathbbm{P}}
\renewcommand{\Pr}{\mathbbm{P}}
\newcommand{\E}{\mathbbm{E}}

\newcommand{\rr}{\mathbbm{R}}
\newcommand{\PT}{\mathcal{P}_T}
\newcommand{\PTp}{\mathcal{P}_{T'}}
\newcommand{\R}{\mathcal{R}}
\newcommand{\polylog}{\mathrm{polylog}}

\newcommand{\set}[1]{\lbrace #1 \rbrace}
\newcommand{\norm}[1]{\lVert #1 \rVert}
\newcommand{\bra}[1]{\langle #1 |}
\newcommand{\ket}[1]{| #1 \rangle}
\newcommand{\braket}[2]{\langle #1 | #2 \rangle}
\newcommand{\calA}{\mathcal{A}}
\newcommand{\CC}{\mathbbm{C}}
\newcommand{\RR}{\mathbbm{R}}
\newcommand{\dg}{\dagger}

\begin{document}

\unitlength=1mm

\title{\Large Continuous-variable quantum compressed sensing}
 
\date{}

\author[1,2]{M.\ Ohliger}
\author[1,3]{V.\ Nesme}
\author[4]{D.\ Gross}
\author[5]{Y.-K.\ Liu}
\author[1]{J.\ Eisert}

\affil[1]{Dahlem Center for Complex Quantum Systems, 
Physics Department, 
Freie Universit{\"a}t Berlin, 14195 Berlin, Germany}
\affil[2]{Institute for Physics and Astronomy, University of Potsdam, 14476 Potsdam, Germany}
\affil[3]{Laboratoire d'Informatique de Grenoble, 38400 Saint-Martin-d'Heres, France}
\affil[4]{Institute of Physics, University of Freiburg, 79104 Freiburg, Germany}
\affil[5]{Applied and Computational Mathematics Division, National Institute of Standards and Technology, Gaithersburg, MD, USA}

\maketitle

\begin{abstract}
We significantly extend recently developed methods to faithfully reconstruct unknown quantum states that are approximately low-rank,
using only a few measurement settings. Our new method is general enough to allow for measurements from a continuous family, and 
is also applicable to continuous-variable states. As a technical result, this work generalizes quantum compressed sensing to the situation where the measured observables are taken from a so-called tight frame (rather than an orthonormal basis) --- hence covering
most realistic measurement scenarios. 
As an application, we discuss the reconstruction of quantum states of light from homodyne detection and other types of measurements, and we present simulations that show the advantage of the proposed compressed sensing technique over present methods. 
Finally, we introduce a method to construct a certificate which guarantees the success of the reconstruction with no assumption on the state, and we show how slightly more measurements give rise to ``universal'' state reconstruction that is highly robust to noise. 
\end{abstract}

\section{Introduction}
One of the most fundamental tasks in quantum mechanics is that of quantum state tomography, i.e.,
reliably reconstructing an unknown quantum state from measurements. Specifically in the context of quantum information processing in most experiments one has to eventually show what state had actually been prepared. Yet, surprisingly little attention has so far been devoted
to the observation that standard methods of quantum state tomography scale very badly with the system size. Only
quite recently, novel more efficient methods have been introduced which solve this problem in a more favorable way in the number of measurement settings that need to be performed \cite{davidletter,david,Kosut,MPS,Seth,yikai,poulin}. 
This development is more timely than ever, given that the experimental progress with 
controlled quantum systems such as trapped ions is so rapid that 
traditional methods of state reconstruction will fail: E.g., $14$ ions can already be controlled in their quantum state \cite{Ions}.
Hence, further experimental progress appears severely challenged as long as ideas of reconstruction cannot keep up. 
Such new methods are based on ideas of quantum compressed
sensing \cite{davidletter,david,yikai} --- inspired by recent work on low-rank matrix completion \cite{candesRecht,candesTao} --- 
or on ideas of approximating unknown quantum states with matrix-product states \cite{MPS}.
Indeed, using methods of quantum compressed sensing, one can reduce the number of measurement settings
from $n^2-1$ in standard methods to $O(rn\log^2 n)$ for a quantum system with Hilbert space dimension $n$,
if the state is of rank $r$. This is efficient in the sense that the number of measurements required is only slightly greater (by an $O(\log^2 n)$ factor) than the number of degrees of freedom in the unknown state.

These ideas have so far been tailored to the situation where observables are taken from an orthonormal operator basis, which is not always the natural situation at hand. 
In this paper, we introduce a theory of state reconstruction based on quantum compressed sensing that 
allows for continuous families of measurements, referred to as {\it tight frames}, 
which can be thought of over-complete, non-orthogonal generalization of operator bases.
These settings are particularly important in the context of continuous-variables, which are notably used to describe quantum optical systems 
beyond the single-photon regime. These have drawn a considerable amount of research, both experimentally and theoretically, due to very desired features such as easy preparation and highly efficient detection. 
Note that when talking about a measurement, we always mean the estimation of an expectation value of an observable for which, of course, several repetition of some experimental procedure are necessary. In this paper we are mainly concerned with the number of distinct observables or \textit{measurement settings} that are needed for tomography.\footnote{Other work 
addresses the number of copies of the unknown state that must be provided \cite{samplecomplexity}
--- that is, the \textit{sample complexity} of tomography.}

In this work, we make significant progress towards a full theory of efficient 
state reconstruction via compressed sensing:
\begin{enumerate}
\item We introduce {\it new incoherence properties} for tight frames, that are sufficient to ensure efficient compressed sensing for low-rank states. This uses an extension of the ``golfing'' proof technique of \cite{davidletter,david}. We give examples of tight frames that satisfy these properties. 
In addition, we show that, if one only wishes to reconstruct ``typical'' or ``generic'' low-rank states, there is a much larger class of tight frames that also lead to efficient compressed sensing.

\item We also describe a way to {\it certify} a successful reconstruction of the state, making our protocol unconditional and heralded. In this way, one does not need to make any
a priori assumptions on the unknown state. 
Our method uses convex duality, and is different from other approaches to certification that focus mainly on pure states \cite{davidletter,MPS,dfe,poulin}.
Also, we discuss the {\it robustness} of the procedure under 
decoherence, imperfect measurements, and statistical noise. We show that, as long as all those effects are small, it is possible to certify that the reconstructed state is close to the true state.

\item We show that, using an incoherent tight frame, and a slightly larger number of measurements, one can achieve \textit{universal} state reconstruction: a single fixed set of measurements can simultaneously distinguish among all possible low-rank states. This is a qualitatively stronger claim than those shown above, and it is obtained using a different technique, based on the ``restricted isometry property'' (RIP) \cite{yikai, cpTightOracleIneq}. This implies strong error bounds, showing that our procedure for state reconstruction is robust to statistical noise, and that it works even when the true state is full-rank with rapidly decaying eigenvalues (in which case our procedure returns a low-rank approximation to the true state).\footnote{As a side note, the RIP-based analysis also shows that the compressed sensing state estimator is nearly minimax-optimal \cite{cpTightOracleIneq}, and it implies nearly-optimal bounds on the sample complexity of low-rank quantum state tomography \cite{samplecomplexity}.}

\item We show how our theory can be applied in realistic experimental scenarios, involving pointwise measurements of the Wigner function, and homodyne detection.

\item We demonstrate {\it numerically} that compressed sensing outperforms the naive approach to tomography not only in the asymptotic limit of large systems but also for system sizes commonly accessible in present day experiments.
\end{enumerate}

This article is organized as follows: We start by introducing quantum compressed sensing in the general setting described by tight frames in Section 2. After discussing a suitable notion of efficiency, we show in Section 3 that efficient compressed sensing is possible if the tight frame fulfills certain incoherence properties. Section 4 is devoted to certified compressed sensing. We discuss how to certify the success of the reconstruction without prior assumptions on the tight frame, both in the ideal case and under the effects of errors. In Section 5 we show universal state reconstruction and error bounds. In Section 6, we investigate applications of the formalism to two common classes of quantum optical experiments; and in Section 7, numerical data, showing the efficiency for small systems, is presented.

\section{Quantum compressed sensing}
Consider a quantum system with Hilbert space dimension $n$. In most cases of interest, $n$ is very large, but the states one wants to reconstruct are approximately low-rank, that is, they are well-approximated by density matrices having rank $r \ll n$. (Pure states correspond to the case where $r = 1$.) When dealing with continuous-variable systems, we will truncate the infinite-dimensional Hilbert space and choose $n$ to be some large but finite cutoff. This is unavoidable, if one wants to do tomography as one cannot reconstruct a state that contains an infinite number of completely independent parameters. However, in most experimentally relevant situations, e.g., continuous-variable light modes with finite mean energy, all states can be arbitrarily well approximated by finite-dimensional ones. We will elaborate on this claim when discussing other sources of errors such as decoherence or imperfect measurements. 

Compressed sensing contains two key ideas. First, rather than measuring all $n^2$ degrees of freedom, it is sufficient to measure a randomly chosen subset of about $rn$ degrees of freedom, provided these degrees of freedom satisfy certain \textit{incoherence properties}. Secondly, one can reconstruct the state using an efficient algorithm. The obvious approach of searching for the lowest-rank state compatible with the measurement results leads to a computationally intractable problem (generally NP-hard). Instead, one can perform a \textit{convex relaxation}, 
and minimize $\|.\|_1$ instead of the rank. Here $\|.\|_p$ stands for the Schatten $p$-norm: $\|.\|_1$, $\|.\|_2$, and $\|.\|=\|.\|_{\infty}$ are respectively the trace norm, Frobenius norm, and spectral (or operator) norm.

Let us denote the $m$ measured observables, i.e.\ Hermitian matrices, by $w_1,\ldots,w_m$, and suppose that we estimate their expectation values (by measuring many copies of the unknown state). Knowing these expectation values (for an unknown state $\rho$) is equivalent to knowing the value of the \textit{sampling operator} $\mathcal{R}(\rho)$, where we define 
\begin{equation}
\label{eq:Rdef}
	\mathcal{R}:\sigma\mapsto\frac{n^2}{m}\sum_{i=1}^m(w_i,\sigma)w_i
\end{equation}
where $(A,B)={\rm Tr}(A^\dag B)$ is the Hilbert-Schmidt scalar product. 
In all of our compressed sensing schemes, $w_1,\ldots,w_m$ will be chosen independently at random from some distribution $\mu$. The sampling operator $\mathcal{R}$ is a linear super-operator on the $d^2$-dimensional real vector space of Hermitian matrices, or operators, $\mathcal{B}(\mathbbm{C}^n)$. Such super-operators will always be denoted by capital script letters. Sometimes we will use the notation $\mathcal{R}\sigma$, multiplying the ``matrix'' $\mathcal{R}$ by the ``vector'' $\sigma$; this means the same thing as $\mathcal{R}(\sigma)$.

Let $\rho$ be the unknown state. In the ideal case, with perfect measurements and no statistical noise, we measure $\mathcal{R}(\rho)$ exactly. Then the procedure to reconstruct $\rho$ can be written as 
\begin{equation}
\min_{\sigma\in\mathcal{B}(\mathbbm{C}^n)} \|\sigma\|_1\,,\quad{\rm subject\,\,to}\,\,\mathcal{R}(\sigma)=\mathcal{R}(\rho).
\label{eq:reconstruction}
\end{equation}
Note that a quantum state $\rho$ is a Hermitian matrix with the additional properties $\rho\ge 0$ and $\Tr\,\rho=1$. However, the reconstruction procedure (\ref{eq:reconstruction}) does not make use of this property and is, therefore, also applicable in more general settings, e.g.\ matrix completion. This problem can be stated as a semi-definite program (SDP) and, therefore, solved efficiently with many well-developed tools. 

In the case of noisy data, we measure $\mathcal{R}(\rho)$ approximately, that is, we measure some $b$ such that $\|b - \mathcal{R}(\rho)\| \leq \delta$, for some norm $\| \cdot \|$ and tolerance $\delta$ that are chosen depending on the kind of noise that is expected. The constraint $\mathcal{R}\sigma=\mathcal{R}\rho$ in (\ref{eq:reconstruction}) can then be replaced by $\| \mathcal{R}(\sigma) - b \| \leq \delta$, which implies $\| \mathcal{R}(\sigma-\rho) \| \leq 2\delta$. 

We remark that equation (\ref{eq:reconstruction}) is the key to certifying our estimate for $\rho$. Notice that if the solution $\sigma^*$ of (\ref{eq:reconstruction}) is unique and fulfills $\|\sigma^*\|_1=1$, then it must be the case that $\sigma^* = \rho$. We will show later on how one can test the uniqueness of the solution $\sigma^*$, without assuming anything about $\rho$. (This can be adapted to work with noisy data, without assuming anything about the noise.)

\subsection{Measurements and tight frames}
When we talk about a compressed sensing scheme, we mean any protocol based on the reconstruction procedure (\ref{eq:reconstruction}), with \textit{some} choice of measurements described by the sampling operator (\ref{eq:Rdef}). In Refs.~\cite{davidletter,david}, the observables were required to be chosen uniformly at random from an operator basis. We substantially generalize these techniques, using the notion of a tight frame, which naturally captures many useful quantum measurements:

\begin{definition}[Tight frame] Let $\mu$ be a probability measure on some set $S$, and for every 
\label{def:tfcondition}
$\alpha\in S$, let $w_\alpha$ be an observable, i.e., a Hermitian operator, and let $\mathcal{P}_{\alpha}$ be the (unnormalized) orthogonal projector which acts as $\mathcal{P}_{\alpha}:\sigma\mapsto (w_\alpha,\sigma)w_\alpha$. We say that $(w_\alpha)_{\alpha\in S}$ is a tight frame if
\begin{equation}
\label{eq:tfcondition}
\int \mathcal{P}_{\alpha} {\rm d}	\mu(\alpha) =\frac{\id}{n^2}.
\end{equation}
\end{definition}
This can also be written as $\EE_\alpha (n^2\mathcal{P}_{\alpha})=\id$ where $\alpha$ is drawn according to $\mu$. Because we deal with randomly drawn operators very often, $\alpha$ will usually denote a random element of $S$ that has distribution $\mu$. 
Note that we do not require that $\|w_\alpha\|_2 = 1$ for all $\alpha$ as it will be convenient in many applications. However, we do require a weaker normalization condition: $\EE_\alpha [ \|w_\alpha\|_2^2 ] = 1$ which follows by taking the trace of (\ref{eq:tfcondition}).

We also define a generalized notion of a tight frame, where the sampling operator is not a sum of projectors; we will need this later to model homodyne detection on optical modes, where a single measurement setting provides more information than only one expectation value.
\begin{definition}[Generalized tight frame]
\label{def:tfcondition2}
Let $\mu$ be a probability measure on some set $S$, and for every $\alpha\in S$ let $\mathcal{Q}_\alpha$ be a positive operator. We say that $(\mathcal{Q}_\alpha)_{\alpha\in S}$ forms a generalized tight frame if
\begin{equation}
\label{eq:tfcondition2}
\int \mathcal{Q}_{\alpha} {\rm d}\mu(\alpha) =\frac{\id}{n^2}.
\end{equation}
\end{definition}
We note that the formalism can be also applied to 8-port homodyne detection which corresponds, for a single mode, to projections on coherent states $|\alpha\rangle$ with $\alpha\in\mathbbm{C}$.

\subsection{Uniqueness of the solution to (\ref{eq:reconstruction})}

For $\rho$ to be the unique solution to (\ref{eq:reconstruction}), any deviation $\Delta$ must be either 
trace-norm increasing, i.e., 
$\|\rho+\Delta\|_1>\|\rho\|_1$, or infeasible, i.e., 
$\mathcal{R}\Delta\neq 0$. This is done by decomposing $\Delta$ into a sum $\Delta_T+\Delta_T^\perp$, and then showing that, with high probability, in the case where $\Delta_T$ is large, $\Delta$ must be infeasible, while in the case where $\Delta_T$ is small, $\Delta$ must be trace-norm increasing. 
Here, we denote by $T$ the real space of Hermitian matrices that send the 
kernel of $\rho$ on its image. In other words, the elements of $T$ are the Hermitian matrices $\sigma$ whose restriction on and to the kernel of $\rho$, i.e. $\pi\sigma \pi$ where $\pi$ is the orthogonal projection on $\operatorname{Ker}\rho$, is equal to $0$. $\mathcal{P}_T$ denotes the projection on this space $T$. 
	
Again, in the actual reconstruction, no assumptions
have to be made concerning $\rho$ or $T$. 
Theorem~\ref{lem:certified}
gives a sufficient condition for uniqueness. The sign function ${\rm sgn}$ of a Hermitian matrix is defined by applying the ordinary sign function to the matrix' eigenvalues. 
\begin{theorem}[Uniqueness of the solution]
\label{lem:certified}
Let $Y\in{\rm range}\,\mathcal{R}$, and set (a) $c_1:=\|\PT Y-{\rm sgn}\,\rho\|_2$, (b) 
$c_2:=\|\PT^\perp Y\|$, and (c) $c_3:=\|\PT\R\PT-\PT\|$. If 
\begin{equation}
	\label{eq:certified}
	\frac{1}{n^3}(1-c_2)\sqrt{\frac{1-c_3}{m}}-c_1>0, 
\end{equation}
then the solution to (\ref{eq:reconstruction}) is unique.
\end{theorem}
Proof: $\Delta$ must be infeasible if $\|\R\Delta\|>0$ which is the case if
\begin{equation}
\label{eq:temp11}
\|\R\Delta_T\|^2_2 = (\R\Delta_T,\R\Delta_T)
>\|\R\Delta_T^\perp\|^2_2.
\end{equation}
The right-hand side is bounded as $\|\R\Delta_T^\perp\|^2_2\le\|\R\|^2\|\Delta_T^\perp\|_2^2\le n^8\|\Delta_T^\perp\|_2^2$ while the left-hand side fulfills 
\begin{align}
\|\R\Delta_T\|_2^2=&(\R\Delta_T,\R\Delta_T)\ge\frac{n^2}{m}(\Delta_T,\R\Delta_T)\nonumber\\
\ge&\frac{n^2}{m}\left(1-\|\PT\R\PT-\PT\|\right)\|\Delta_T\|_2^2.
\end{align}
Thus, (\ref{eq:temp11}) is satisfied if 
 \begin{equation}
 \frac{n^2}{m}\left(1-\|\PT\R\PT-\PT\|\right)\,\|\Delta_T\|^2_2>n^8\|\Delta_T^\perp\|_2^2, 
\end{equation}
which, using definition (c), is equivalent to 
\begin{equation}
\label{eq:th1d}
\|\Delta_T^\perp\|_2<\frac{1}{n^3}\|\Delta_T\|_2\sqrt{\frac{1-c_3}{m}}. 
\end{equation}
Using the pinching \cite{bhatia} and H{\"o}lder's inequalities, as detailed in Ref.~\cite{david}, yields 
\begin{equation}
\label{eq:pinching}
\|\rho+\Delta\|_1\ge\|\rho\|_1+({\rm sgn}\,\rho+{\rm sgn}\,\Delta_T^\perp,\Delta).
\end{equation}
The second term is equal to
\begin{equation}\label{eq:pinching2}
({\rm sgn}\,\rho-Y,\Delta_T)+({\rm sgn}\,\Delta_T^\perp-Y,\Delta_T^\perp)
\end{equation}
which is, according to (a) and (b), larger than 
\begin{equation}
\|\Delta_T^\perp\|_2-c_2\|\Delta_T^\perp\|_2-c_1\|\Delta_T\|_2.
\end{equation}
Inserting this into (\ref{eq:th1d}) gives rise to condition (\ref{eq:certified}) and concludes the proof. If all the elements in the tight frame fulfill $\|w_\alpha\|_2=1$ we call it normalized and one can bound $\|\mathcal{R}\|\le n^2$. In this case (\ref{eq:certified}) in Theorem \ref{lem:certified} can be weakened to
\begin{equation}
	\label{eq:certifiedweaker}
	\frac{1}{n}(1-c_2)\sqrt{\frac{1-c_3}{m}}-c_1>0.
\end{equation}

\subsection{Efficient quantum compressed sensing}
Let $\rho$ be a state of dimension $n$ and rank $r$. In the compressed sensing method of tomography, we choose $m$ observables $w_1,\ldots,w_m$ randomly from some distribution, measure their expectation values with respect to $\rho$, then solve (\ref{eq:reconstruction}) to obtain $\sigma^*$, which is our estimate of $\rho$. 

For a given state $\rho$, there is some probability $p_f(\rho)$ that the procedure may fail (i.e., it may return a solution $\sigma^*$ that is not close to $\rho$). Note that this probability $p_f(\rho)$ is taken with respect to the random choice of $w_1,\ldots,w_m$, and the random outcomes of the measurements. We say that the method succeeds with high probability if, for every low-rank state $\rho$, the failure probability is $p_f(\rho)$ small. Equivalently, the method succeeds with high probability if, 
\begin{quotation}\noindent
for every low-rank state $\rho$, most choices of the observables $w_1,\ldots,w_m$ can be used to successfully reconstruct $\rho$. 
\end{quotation}

Now, the basic question is: how large does $m$ have to be, to ensure that the method succeeds with high probability? A common situation is that the system under consideration consists of $k$ subsystems with local Hilbert space dimension $d$; then $n=d^k$. Of course, no method of tomography can counter the exponential growth of the required number of measurements in $k$. Thus, efficiency needs to be regarded relative to the $n^2-1$ measurements necessary for standard tomography. As even a pure state needs $\Theta(n)$ parameters to be described, this also is a lower bound to the number of observables that need to be measured. We allow for an additional polylogarithmic overhead and define efficiency as follows: 

\begin{definition}[Efficient quantum compressed sensing]
\label{def:efficient}
Compressed sensing for a state $\rho$ (with dimension $n$ and rank $r$) is regarded as 
efficient if: The number of measured observables satisfies $m=O(nr\,\polylog(n))$, and the probability of failure satisfies $p_f(\rho) < 1/2$. 
\end{definition}
If this is the case, $p_f(\rho)$ can be made arbitrarily small by repeating the protocol and using a majority vote among the reconstructed states to get the final result. Then, the probability of failure decays exponentially in the number of repetitions.

Note that this is a very stringent definition of efficiency. One can also merely ask for any scaling of $m$ in $o(n^2)$. Of course, this weaker condition is easier to satisfy, as we shall see later on.

\subsection{Sufficient conditions for efficiency}

The general theory of quantum
compressed sensing, which will be developed here, 
relies heavily on and significantly extends 
the analysis for the special case where the observables form an operator basis in Ref.\ \cite{david}. The hypothesis for Theorem \ref{lem:certified} is fulfilled if $c_1\le 1/(2n^4)$, $c_2\le 1/2$, $c_3\le 1/2$ under the additional condition $m<n^2/2$, which can be safely assumed to be true as we are only interested in the regime of $m\ll n^2$. For normalized tight frames, the first condition can be weakened to $c_1\le 1/(2n^2)$. We show conditions to the tight frame under which those conditions are fulfilled with high probability. 

For efficient compressed sensing to be possible, the observables $w_\alpha$ need to fulfill certain \textit{incoherence properties}. Roughly speaking, the observables are ``incoherent'' if they have small inner product with every possible state one wishes to reconstruct. For example, operator norm can be a measure of incoherence for reconstructing pure states, since $\|w_\alpha\| = \max_{\braket{\psi}{\psi}=1} \bra{\psi}w_\alpha\ket{\psi}$. We distinguish two general cases (which we will define more precisely in the following sections): 
\begin{enumerate}
\item ``Fourier-type'' compressed sensing, where almost all of the observables have small operator norm. In this case, efficient compressed sensing is possible for any low-rank state.
\item ``Non-Fourier type'' compressed sensing, where the observables may have large operator norm, but efficient compressed sensing is still possible for certain restricted classes of states, e.g., generic states.
\end{enumerate}

\subsection{Fourier-type efficient compressed sensing}
The efficiency of a tomography protocol, as given in Definition \ref{def:efficient}, is a statement about a family of procedures acting on systems with growing dimension $n$. We now give a sufficient condition for a family of tight frames to allow for efficient compressed sensing. 

\begin{theorem}[Fourier type]
\label{obs:almostf}
Let $(w_\alpha(n))_{\alpha\in S}$ be, for any $n>0$, a tight frame. Let $\rho(n)$ be any state with dimension $n$ and rank $r$. Let $\nu=O({\rm polylog}(n))$. Set $C(n):=\{\alpha\in S : \,\|w_\alpha(n)\|^2>\nu/n\}$ and let $\mu(C(n))$ be the measure of this set. If
\begin{equation}
\label{eq:mu}
\mu(C(n))\le\frac{1}{16\sqrt{r}n^2m},
\end{equation}
efficient compressed sensing is possible for the family of states $\rho$. 
\end{theorem}
Here, the underlying ``incoherence property'' is the bound on the operator norm of the observables, 
\begin{equation}\label{eq:ftincoherence}
\|w_\alpha(n)\|^2 \leq \nu/n, 
\end{equation}
which holds for ``most'' choices of $\alpha$.
If there is no risk of confusion, we will omit the explicit dependencies on $n$.

\subsubsection{Perfect Fourier-type case}
We have to first consider the case $\mu(C)=0$. Even though the proof in Ref.\ \cite{david} can be applied with only minor changes, we state it in a way as complete and still non-technical as possible where we focus on the asymptotic behavior and do not provide explicit constants. We need Lemma 5 from Ref.\ \cite{david} which reads:
\begin{lemma}[Large deviation bound for the projected sampling operator]
\label{lem:5}
For all $t<2$
\begin{equation}
\Pr\left[\|\PT\mathcal{R}\PT-\PT\|> t\right]\le 4nr\exp\left(-\frac{t^2\kappa}{8\nu}\right),
\end{equation} 
where $\kappa=m/(nr)$ is the oversampling factor which must fulfill $\kappa=O({\rm polylog}(n))$ for efficiency. 
\end{lemma}
The tool to prove Lemma \ref{lem:5} and other bounds of this form is provided by the operator-Bernstein inequality which was first given in Ref.\ \cite{ahlswede} and which we state here without a proof.
\begin{lemma}[Operator-Bernstein inequality]
\label{lem:opbernstein}
Let $(X_i)_{i=1,\ldots,m}$ be i.i.d. Hermitian matrix-valued random variables with zero mean. Suppose there exist constants $V_0$ and $c$ such that $\|\EE(X_i^2)\|\le V_0^2$, $\|X_i\|\le c$ where the latter needs to be true for all realizations of the random variable. Define $A=\sum_iX_i$ and $V=mV_0^2$. Then, for all $t\le 2V/c$
\begin{equation}
\Pr\left[\|A\|>t\right]\le 2n\exp\left(-\frac{t^2}{4V}\right).
\end{equation}
\end{lemma}
The proof of Lemma \ref{lem:5} is given in Ref.\ \cite{david} but we restate it here 
because it is quite instructive. Let $\alpha$ be a random variable taking values in $S$. We define $m$ random variables by $Z_{\alpha_i}=(n^2/m)\PT\mathcal{P}_{\alpha_i}\PT$ and $X_{\alpha_i}=Z_{\alpha_i}-\EE(Z_{\alpha_i})$. Now $S=\PT\mathcal{R}\PT-\PT=\sum_i X_{\alpha_i}$ and we have to estimate the maximum of $\|X_{\alpha_i}\|$ and the norm of the variance of $X_{\alpha_i}$ in order to apply Lemma \ref{lem:opbernstein}. From the incoherence condition (\ref{eq:ftincoherence}), we get by using the matrix H{\"o}lder inequality \cite{bhatia} 
\begin{equation}
\|\mathcal{P}_Tw_\alpha\|_2^2=\sup_{\sigma\in T,\|\sigma\|_2=1}(w_\alpha,\sigma)^2\le 2\nu\frac{r}{n}.
\end{equation}
This allows us to write
\begin{align}
\|\EE(X_{\alpha_i}^2)\|=&\|\EE(Z_{\alpha_i}^2)-\EE(Z_{\alpha_i})^2\|\nonumber\\
\le&\frac{2n\nu r-1}{m^2}\|\PT\|\le\frac{2\nu}{m\kappa}\label{eq:V01}
\end{align}
and
\begin{align}
\|X_{\alpha_i}\|=&\frac{1}{m}\|n^2\PT\mathcal{P}_{\alpha_i}\PT-\PT\|\nonumber\\
\le&\frac{1}{m}\|n^2\PT\mathcal{P}_{\alpha_i}\PT\|=\frac{n^2}{m}\|\PT w_{\alpha_i}\|_2^2\nonumber\\
\le&\frac{2\nu}{\kappa}.\label{eq:c1}
\end{align}
Here, and in the remainder, statements of the form (\ref{eq:c1}) are meant to hold for all realization of the random variable as needed in the Operator Bernstein inequality. Inserting now (\ref{eq:V01}) and (\ref{eq:c1}) into Lemma \ref{lem:opbernstein} yields Lemma \ref{lem:5} which
concludes the proof. Applying Lemma \ref{lem:5} for $t=1/2$ and choosing $\kappa=O(\polylog(n))$, the probability that $c_3>1/2$ can be made arbitrarily small. 

Now we have to construct a certificate $Y$ whose projection on $T$ is close to ${\rm sgn}\,\rho$. This is done by an iterative process, called the golfing scheme \cite{david}. The $m$ samples are grouped into $l$ groups which are indexed by $i$ and contain $m_i$ samples each. Let $\mathcal{R}_i$ be the sampling operator of the $i$th group and set $X_0={\rm sgn}\,\rho$, $X_i=(\id-\PT\mathcal{R}_i\PT)X_{i-1}$, $Y_i=\sum_{j=1}^i\mathcal{R}_jX_{j-1}$, and $Y=Y_l$. 

Again, Lemma \ref{lem:5} can be used to show that with high probability (at the expense of a polylog growth of $\kappa_i$)
\begin{equation}
\|X_i\|_2\le\|\PT\mathcal{R}_i\PT-\PT\|\|X_{i-1}\|_2\le\frac{1}{2}\|X_{i-1}\|_2,
\end{equation}
and, therefore, $\|X_i\|_2\le\sqrt{r}2^{-i}$ from which we get
\begin{equation}
c_1=\|X_l\|_2\le\sqrt{r}2^{-l}\le \frac{1}{2n^2},
\end{equation}
while for the final inequality to hold it is enough to set $l=\Theta(\log(\sqrt{r}n))$. For the last remaining condition we need the subsequent statement:
\begin{lemma}[Bound for the orthogonal projection]
Let $F\in T$ and $t\le\sqrt{2/r}\|F\|_2^2$. Then
\begin{equation}
\Pr\left[\|\PT^\perp\mathcal{R}F\|>t\right]\le 2n\exp\left(-\frac{t^2\kappa r}{4\nu\|F\|_2^2}\right).
\end{equation}
\label{lem:7}
\end{lemma}
Proof: Without loss of generality, consider $\|F\|_2=1$ and define the zero-mean random variables $X_{\alpha_i}=(n^2/m)\PT^\perp w_{\alpha_i}(w_{\alpha_i},F)$ which fulfill $\sum_iX_{\alpha_i}=\PT^\perp\mathcal{R} F$. Their variance is bounded by
\begin{align}
	\|\EE(X_{\alpha_i}^2)\|\le&\frac{n^4}{m^2}\int{\rm d}(\mu)\,(w_\alpha,F)^2\|(\PT^\perp w_\alpha)^2\|\nonumber\\	
	\le&\frac{\nu}{m\kappa r},\label{eq:V02}
\end{align}
and their norm by
\begin{equation}
\label{eq:c21}
\|X_{\alpha_i}\|\le\frac{n^2}{m}\sqrt{\frac{\nu}{n}\frac{2\nu r}{n}}=\frac{\sqrt{2}\nu}{\sqrt{r}\kappa}.
\end{equation}
Lemma \ref{lem:7} follows directly from using (\ref{eq:V02}) and (\ref{eq:c21}) in Lemma \ref{lem:opbernstein}. Now we can bound
\begin{equation}
\label{eq:c2}
c_2=\|\PT^\perp Y\|\le\frac{1}{4}\sum_{i=1}^l2^{-(i-1)}<\frac{1}{2}.
\end{equation}
Again, the probability of (\ref{eq:c2}) not being true can be made as small as desired by choosing $\kappa=O(rm\,\polylog(n))$. Of course, this is also true for the total probability of failure which concludes the proof. 

\subsubsection{Imperfect Fourier-type case}

We now show that the incoherence condition may be violated for some of the observables and adapt a technique used in Ref.\ \cite{ripless}. Intuitively, when $\mu(C)$ is small enough, we can just abort and restart the reconstruction procedure whenever we encounter a non-incoherent operator during our sampling process. The probability of this to happen is upper bounded by $(16\sqrt{r}n^2)^{-1}$ as obtained from (\ref{eq:mu}) by a union bound over the $m$ measurements. This is equivalent to sampling only from the set $S\setminus C$. The conditional probability distribution on the observables does fulfill the approximate tight-frame condition 
\begin{equation}
	\label{eq:approxtf}
	\|\mathcal{W}-\id\|\le1/(8\sqrt{r}),
\end{equation}
where $\mathcal{W}=n^2\EE(\mathcal{P}_{\alpha}|E)$ where $E$ is the event that all of the $m$ chosen operators satisfy $\|w_{\alpha_i}\|^2\le\nu/n$ and its complement is denoted by $E^c$. Let $\id_E$ be the indicator function of $E$. Then,
$\id=n^2\EE(\mathcal{P}_\alpha)=n^2\EE(\mathcal{P}_\alpha\id_E)+n^2\EE(\mathcal{P}_\alpha\id_{E^c}).$
This leads to
\begin{multline}
\|n^2\EE(\mathcal{P}_\alpha|E)-\id\|\PP(E)=\|(1-\PP(E))\id-n^2\EE(\mathcal{P}_\alpha\id_{E^c})\|\\
\le\PP(E^c)+n^2\|\EE(\mathcal{P}_\alpha\id_{E^c})\|.
\label{eq:snd2}
\end{multline}
With the help of Jensen's inequality, we can simplify $\|\EE(\mathcal{P}_\alpha\id_{E^c})\|\le\EE(\|\mathcal{P}_\alpha\|\id_{E^c})=\PP(E^c)$. Inserting this into (\ref{eq:snd2}) and rearranging, we get
\begin{equation}
\|n^2\EE(\mathcal{P}_\alpha|E)-\id\|\le\frac{2n^2\PP(E^c)}{1-\PP(E^c)}\le 2n^2\PP(E^c).
\label{eq:snd3}
\end{equation}
Our claim follows by taking $\PP(E^c)=1/(16\sqrt{r})$ which is always true by a union bound. We now have to justify why the tight frame condition (\ref{eq:tfcondition}) can be replaced by the approximate one in Eq.\ (\ref{eq:approxtf}) in the proof of Lemma \ref{lem:5} and Lemma \ref{lem:7}. We denote the probability measure which is conditioned on the event $E$ by $\bar{\mu}$. 

Lemma \ref{lem:5} provides a bound to
\begin{eqnarray}
	\label{eq:lem51}
	\|\PT\left(\mathcal{R}-\id\right)\PT\|
	&\le& \|\PT\left(\mathcal{R}-\mathcal{W}\right)\PT\|\\
	&+&
	\|\PT\left(\mathcal{W}-\id\right)\PT\| .
	\nonumber
\end{eqnarray}
We define the random variables $Z_{\alpha_i}$ and $X_{\alpha_i}$ as in the proof of Lemma \ref{lem:5} and bound their variance as
\begin{eqnarray}
	\label{eq:lem52}
	\|\E(X_{\alpha_i}^2)\|&=&\|\E(Z_{\alpha_i}^2)-\E(Z_{\alpha_i})^2\|\nonumber\\
	&\le&\|\E(Z_{\alpha_i}^2)\|+\|\E(Z_{\alpha_i})^2\|\nonumber\\
	&\le&\frac{1}{m^2}\left(2n\nu r+\|\mathcal{W}\|^2\right)\nonumber\\
	&=&\frac{1}{m^2}\left(2n\nu r+(\frac{1}{8\sqrt{r}}+1)^2\right)\le\frac{4n\nu r}{m^2},
\end{eqnarray}
and their norm as $\|X_{\alpha_i}\|\le 2\nu nr/m$. Using the operator Bernstein inequality yields 
\begin{lemma}[Large deviation bound for the projected sampling operator]
\label{lem:5replace}
\begin{equation}
\PP(\|\PT\mathcal{R}\PT-\PT\|>t)\le 4nr\exp\left(-\frac{t^2\kappa}{64\nu}\right),
\label{eq:lem5new}
\end{equation}
for all $1/(4\sqrt{r})\le t\le 4$.
\end{lemma} 
where we have also used (\ref{eq:lem51}) to bound the second term in (\ref{eq:approxtf}). Thus, up to an irrelevant constant factor, Lemma \ref{lem:5replace} replaces Lemma \ref{lem:5} wherever it is used. 

To also replace Lemma~\ref{lem:7}, let $F\in T$, $\|F\|_2=1$ and note that
\begin{equation}
	\|\PT^\perp\R F\|\le\|\PT^\perp(\R-\mathcal{W})F\|+\frac{1}{8\sqrt{r}}. 
\end{equation}	
The random variables are $Z_{\alpha_i}=(n^2/m)\PT^\perp\mathcal{P}_{\alpha_i}F$ and $X_{\alpha_i}=Z_{\alpha_i}-\E(Z_{\alpha_i})$ where the variance is bounded by
\begin{eqnarray}
	\|\E(X_{\alpha_i}^2)\|&=&\|\E(Z_{\alpha_i}^2)-\E(Z_{\alpha_i})^2\|\nonumber\\
	&\le&\|\E(Z_{\alpha_i}^2)\|+\|\E(Z_{\alpha_i})^2\|\nonumber\\
	&\le&\frac{1}{m^2}\left(n\nu+\frac{1}{64r}\right)\le\frac{2\nu}{m\kappa r}
\label{eq:lem72}
\end{eqnarray}
which gives, together with $\|X_{\alpha_i}\|\le 2\sqrt{2}\nu/(\sqrt{r}\kappa)$, and an application of the operator-Bernstein inequality the subsequent statement.

\begin{lemma}[Bound for the orthogonal projection]
\label{lem:7replace}
Let $F\in T$ and $1/(2\sqrt{r})\le t/\|F\|_2\le 2\sqrt{2/r}$. Then
\begin{equation}
	\label{eq:lem7}
	\Pr\left[\|\PT^\perp F\|>t\right]\le 2n\exp\left(-\frac{t^2\kappa r}{32\nu\|F\|_2^2}\right).
\end{equation}
\end{lemma} 
Lemma \ref{lem:7replace} takes the place of Lemma \ref{lem:7} and, again, differs only by a constant factor in the exponent which concludes the proof of Theorem \ref{obs:almostf}.

An example for a Fourier-type frame for which $\mu(C)\neq 0$ is given by the following situation.
Here, with some probability, every Hermitian matrix with unit Frobenius norm is drawn in the measurement.
\begin{example}[Tight frame containing all Hermitian matrices]
Any $w_\alpha\in\mathcal{B}(\mathbbm{C}^n)$ with $\|w_\alpha\|_2=1$ can be viewed as a vector on the $n^2$ dimensional unit sphere. Therefore, on can define a rotationally invariant Haar measure on it. The tight frame formed by the Haar measure on all Hermitian matrices with $\|w_\alpha\|_2=1$ fulfills Theorem \ref{obs:almostf}. Therefore, it allows for efficient compressed sensing.
\end{example}
In order to satisfy~Theorem \ref{obs:almostf}, we have to show
\begin{equation}
	\label{eq:app1}
	\PP\left(\|w_\alpha\|^2>\frac{\nu}{n}\right)\le\frac{1}{16\sqrt{r}n^2m},
\end{equation}
where $\nu=O\left({\rm polylog}\,(n)\right)$. To see that this is true, we note that we are dealing with a normalized version of the extensively discussed Gaussian unitary ensemble (GUE) 
denoted by 
$\{\bar{w}_\alpha\}$, $w_\alpha=\bar{w}_\alpha/\|w_\alpha\|_2$. 
Now for all $\delta>0,\varepsilon>0$ we have
\begin{equation}
\PP\left(\|w_\alpha\|\ge\frac{\delta}{\sqrt{n}}\right)\le\PP\left(\|\bar{w}_\alpha\|>\frac{\delta\varepsilon}{\sqrt{n}}\right)+\PP\left(\|\bar{w}_\alpha\|_2>\varepsilon\right).
\end{equation} 
The first term can be bounded using a result from Ref.\ \cite{randommatrix} yielding
\begin{equation}
	\PP(\|\bar{w}_\alpha\|>{\delta\varepsilon}/{\sqrt{n}})\le c_1\exp(-c_2n(\delta\varepsilon-2)^{3/2})
\end{equation}
where $c_1,c_2>0$ 
are small constants while for the second term we use the properties of the $\chi_k^2$-distribution which are given the appendix. From this, we get 
\begin{equation}
	\PP\left(\|\bar{w}_\alpha\|^2_2>1-y\right)\le\exp(-y^2n^3/4).
\end{equation}
We set $y=1/2$ and see that (\ref{eq:app1}) is fulfilled for some constant $\nu$ when $n$ is large enough.

Product measurements are of great experimental importance: They describe the situation
of addressing individual quantum systems, say, ions in an ion trap experiment or 
individual modes in an optical one. They 
are described by tight frames which are formed as tensor products of tight frames on the local systems. Given a tight frame which fulfills $\|w_\alpha\|^2\le \nu/d$, one can obtain a tight frame on the $n=d^k$ dimensional Hilbert space by forming the $k$-fold tensor product. The strongest possible incoherence property we can obtain is $\|w_{\alpha}\|^2\le\nu^k/n$. Unless $\nu=1$, as for the Pauli matrices, $\nu$ grows too fast to allow for efficient compressed sensing for all states. This is even true if the incoherence condition may be violated on some set $C$ with $\mu(C)=O(1/{\rm poly}(n))$.

\subsection{Non-Fourier-type efficient compressed sensing}

The conditions in Theorem \ref{obs:almostf} imply that efficient compressed sensing is possible for \textit{any} low-rank state $\rho$. This is a quite special situation and for Theorem \ref{obs:almostf} to be fulfilled, either a very special structure, like the one of the Pauli basis \cite{davidletter}, or a large amount of randomness, like in the above example, is needed. As an example for a very different situation, consider the state $\rho=|0\rangle\langle 0|$ together with the observables which corresponds to the sampling of single matrix-entries (or the Hermitian combinations of two of them). Here, one needs to take $\Theta(n^ 2)$ attempts until one ``hits'' the single non-zero entry in the upper-left corner. This is not surprising because the operators in this basis fulfill $\|w_\alpha\|=\Theta(1)$. However, for most of the states, efficient compressed sensing is indeed possible in this basis. In Theorem \ref{obs:nf}, we give a sufficient condition for combinations of states and tight frames to work.

\begin{theorem}[Non-Fourier-type efficient compressed sensing]
\label{obs:nf}
For a given tight frame $\set{w_\alpha \;|\; \alpha \in S}$, and a given rank-$r$ state $\rho$, denote by $C\subset S$ the set of observables for which at least one of the following conditions is not fulfilled:
\begin{align}
\|\PT w_\alpha\|_2^2\le&\frac{2\nu r}{n},\\
(w_\alpha,{\rm sgn}\,\rho)^2\le&\frac{\nu r}{n^2}.\label{eq:2nd}
\end{align}
If $\mu(C)\le(16\sqrt{r}n^2m)^{-1}$, efficient compressed sensing is possible for the state $\rho$. 
\end{theorem}
The golfing scheme works exactly like in the Fourier-type case, as does the proof of Lemma~\ref{lem:5}. However, Lemma~\ref{lem:7replace} must be replaced by something else. Again, we use the technique of conditioning which means that we assume the incoherence condition to hold for all operators in the tight frame and the tight frame condition to be approximately true as in (\ref{eq:approxtf}). First, we need some preparation.
\begin{lemma}[Bound to the scalar product]
\label{lem:9}
Let $F\in T$ such that $\|F\|_2\le f$, $1/(4\sqrt{r})\le f/t\le 2\sqrt{2/r}$, and 
\begin{equation}
\label{eq:lem9}
(w_\alpha,F)^2\le\frac{\nu f^2}{n^2}
\end{equation}
for all $\alpha\in S$. Then
\begin{equation}
\Pr\left(\|\PT^\perp\mathcal{R}F\|>t\right)\le2n\exp\left(-\frac{t^2\kappa r}{64\nu f^2}\right).
\end{equation}
\end{lemma}
Proof: We consider the same same random variables as in the proof of Lemma \ref{lem:5replace} (note that we have again set $\|F\|_2=1$) and bound their variance as
\begin{align}
\|\EE(X_{\alpha_i}^2)\|\le&\frac{n^4}{m^2}\Bigl(\max_\psi\int{\rm d}\mu(\alpha)\,(w_\alpha,F)^2\langle\psi|w_\alpha^2|\psi\nonumber\rangle+\frac{1}{64r}\Bigr)\\
\le&\frac{4\nu}{m\kappa r}\label{eq:V03},
\end{align}
where we have used the incoherence property and 
\begin{equation}
\label{eq:wa2}
	\left\|\int{\rm d}\bar{\mu}(\alpha)\,w_\alpha^2\right\|\le\frac{2}{n}.
\end{equation}
To see that (\ref{eq:wa2}) holds, we start with
\begin{equation}
\frac{\id}{n}=\int{\rm d}\mu(\alpha)\,w_\alpha^2=(1-|C|)\int{\rm d}\bar{\mu}(\alpha)\,w_\alpha^2+\int_C{\rm d}\mu(\alpha)\,w_\alpha^2
\end{equation}
where the first equality follows directly from the tight frame property, c.f.\ Ref.~\cite{david}, while the second one stems from the definition of the conditional probability distribution $\bar{\mu}$. Rearranging and taking the norm yields
\begin{equation}
\left\|\int{\rm d}\bar{\mu}(\alpha)\,w_\alpha^2\right\|\le\frac{1}{1-|C|}\left(\frac{1}{n}+|C|\right)
\end{equation}
which implies (\ref{eq:wa2}). Using (\ref{eq:V03}) together with $\|X_{\alpha_i}\|\le 2\sqrt{2}\nu/(\sqrt{r}\kappa)$ in Lemma \ref{lem:opbernstein}, we obtain Lemma \ref{lem:9} which concludes the proof.

The above Lemma must by applied for $F=X_0,\ldots,X_l$, i.e., the operators occurring in the golfing scheme. By the second incoherence condition, (\ref{eq:lem9}) is fulfilled for $F=X_0$. To ensure that incoherence is preserved during the golfing scheme, we must use a more complicated and technical argument than in Ref.~\cite{david} where a union bound over all elements of the operator basis was used which is clearly impossible in a tight frame with an infinite number of elements. 

\begin{lemma}[Replacing the union bound]
\label{lem:10}
\begin{equation}
\label{eq:lem10}
\mathbbm{P}_{\mathcal{R}}\left(\xi((\mathbbm{1}-\mathcal{P}_T\mathcal{R}\mathcal{P}_T)F)>\frac{1}{2}\|F\|^2\right)\le 16\sqrt{r}mn^2\exp\left(-\frac{\kappa}{64\xi(F)\nu}\right),
\end{equation}
where $\xi(F)$ 
is the smallest number such that
\begin{equation}
\Pr_{\alpha}\left((w_\alpha,F)^2<\xi(F)\right)\le\frac{1}{16\sqrt{r}n^2m}.
\end{equation}
\end{lemma}
Proof: We fix an element $w_\beta$ from the tight frame and note that for $F\in T$ 
\begin{align}
|(w_\beta,\mathcal{P}_T(\mathcal{R}-\mathbbm{1})F)|\le&|(w_\beta,\mathcal{P}_T(\mathcal{R}-\mathcal{W})F)|\nonumber\\
&+|(w_\beta,\mathcal{P}_T(\mathcal{W}-\mathbbm{1})F)|\label{eq:5}.
\end{align}
The latter term is smaller than $\|\mathcal{W}-\mathbbm{1}\|\|F\|_2$. To bound the former term, we define the random variable 
\begin{equation}
X_{\alpha_i}=\frac{1}{m}(w_\beta,F)-(w_\beta,\frac{n^2}{m}\mathcal{P}_T w_{\alpha_i})(w_{\alpha_i},F)
\end{equation}
whose variance is bounded by
\begin{equation}
\label{eq:varbound}
|\mathbbm{E}[X_{\alpha_i}^2]\le\frac{2n\xi(F)\nu r}{m^2}+\frac{1}{m^2}\|\mathcal{W}-\mathbbm{1}\|^2\|F\|_2^2
\end{equation}
and $\|X_{\alpha_i}\|\le 2(1+n\nu r)\sqrt{\xi(F)}/m$. Using once again the operator Bernstein inequality yields after squaring
\begin{equation}
	\label{eq:lem102}
	\mathbbm{P}\left((w_\beta,(\mathbbm{1}-\mathcal{P}_T\mathcal{R}\mathcal{P}_T)F)^2>	
	\frac{1}{2}\|F\|_2^2\right)
	\le2\exp\left(-\frac{m}{128nr\xi(F)\nu}\right).
\end{equation}
Eq.~(\ref{eq:lem102}) says that the desired property is true with high probability for any fixed $w_\beta$. To show that it is also true with high probability for most of the operators, we need a simple fact from probability theory, which is shown in the appendix.
\begin{lemma}[Inverting probabilities]
\label{lem:probtheory}
Let $X$ and $Y$ be two measure spaces and denote by $x\sim y$ a relation between the elements $x\in X$ and $y\in Y$. If
\begin{equation}
\forall x\in X:\,\mathbbm{P}(x\not\sim y|y\in Y)\le p
\end{equation}
then
\begin{equation}
\mathbbm{P}\left(\mathbbm{P}(x\not\sim y|x\in X)>\beta|y\in Y\right)\le\frac{p}{\beta}
\end{equation}
\end{lemma}
Applying this to (\ref{eq:lem102}) and using the definition of $\xi(F)$, one directly obtains (\ref{eq:lem10}) which completes the proof of Lemma \ref{lem:10}. Now, we can see that $\mu(X_i)\le 2^{-i}\sqrt{r}\nu/n^2$ which means that Lemma \ref{lem:9} can be applied in the golfing scheme and we have proven Theorem \ref{obs:nf}.

\subsection{Reconstructing generic quantum states}

In a next step, we investigate the reconstruction of random quantum states, 
that are sampled from probability measures that are
invariant under the action of the unitary group by conjugation.
We show examples of tight frames that satisfy the incoherence properties required in Theorem \ref{obs:nf} to allow reconstruction of \textit{most} quantum states.

\begin{theorem}[Incoherence properties of generic states]
\label{obs:allgood}
Let $(w_\alpha)_{\alpha\in S}$ be a (family of) tight frame for which all operators fulfill $\|w_\alpha\|_1=O({\rm polylog}(n))$, and pick a random rank $r$ quantum state $\rho$, with a distribution that is invariant under the action of the unitary group. Then the probability that $\rho$ cannot be efficiently reconstructed by compressed sensing vanishes as $O(1/{\rm poly}\,(n))$. 
\end{theorem}
Note that Theorem \ref{obs:allgood} holds for all unitarily invariant measures on the quantum states of rank $r$ regardless of the actual distribution of the eigenvalues.

Proof: We first show that for any fixed element of the tight frames, both incoherence properties are fulfilled with high probability. First, we turn to
\begin{equation}
\label{eq:eq1}
\|P_T w\|_2^2=\sum_{i,j|{\rm min}(i,j)\le r}|(U^\dag w U)_{i,j}|^2
\end{equation}
where $U$ is a unitary matrix which is chosen according to the Haar measure and we have fixed an element $w$ from the tight frame. We look at the $i$th row of $U^\dag w U$ and note that
$\sum_j|(U^\dag w U)_{i,j}|^2=\sum_j|(U^\dag w)_{i,j}|^2$. We write $w_j$ for the $j$th column vector of $w$ and note that $U^\dag w_j/\|w_j\|$ is just a random vector on a sphere . Thus, the squares of its coordinates are concentrated around $1/n$, c.f.\ the appendix, and we get 
\begin{equation}
\mathbbm{P}_U\left(\frac{|(U^\dag w)_{i,j}|^2}{\|w_j\|^2}>\frac{\nu}{n}\right)\le 2\exp\left(-\frac{\nu}{8}\right).
\end{equation}
Using this in Eq.\ (\ref{eq:eq1}), inserting $\sum_i\|w_i\|^2=\|w\|^2_2=1$, and applying a union bound yields
\begin{equation}
\mathbbm{P}_U\left(\|\mathcal{P}_T w\|_2^2>\frac{2\nu r}{n}\right)\le 2nr\exp\left(-\frac{\nu}{8}\right).
\end{equation}
Employing again Lemma \ref{lem:probtheory}, this implies
\begin{multline}
\label{eq:eq2}
\mathbbm{P}_U\left(\mathbbm{P}_w\left(\|\mathcal{P}_Tw\|_2^2>\frac{2\nu r}{n}\right)>\frac{1}{16\sqrt{r}n^2m}\right)
\le 32r^{3/2}n^3m \exp\left(-\frac{\nu}{8}\right).
\end{multline}
where $w$ is chosen according to the probability distribution of the tight frame. 
By allowing $\nu$ to grow polylogarithmically in $n$, this probability vanishes polynomially in $n$ which means that it is violated to much only for a proportion of state vanishing polynomially as $n$ grows. 
Now we turn to the second non-Fourier incoherence condition. Decomposing $w$ as a sum of projectors on orthogonal subspaces $w=\sum_k \lambda_k |\Psi_k\rangle\langle \Psi_k|$, we can write
\begin{equation}
|(w,{\rm sgn}\,\rho)|\le\sum_{i=0}^r\sum_k|\lambda_k||\langle i|U^\dag|\Psi_k\rangle|^2.
\end{equation}
Using the concentration of measure on the sphere and $\sum_k|\lambda_k|=\|w\|_1$ yields
\begin{equation}
\mathbbm{P}\left((w,{\rm sgn}\,\rho)^2>\frac{r^2\nu}{n^2}\|w\|_1^2\right)\le 2nr\,e^{-\sqrt{\nu}/8},
\end{equation}
which finally gives
\begin{multline}
\label{eq:eq3}
\mathbbm{P}_U\left(\mathbbm{P}_w\left((w,{\rm sgn}\,\rho)^2>\frac{r^2\nu}{n^2}\|w\|_1^2\right)>\frac{1}{16\sqrt{r}mn^2}\right)\\
\le 32r^{3/2}mn^3\exp\left(-\frac{\sqrt{\nu}}{8}\right).
\end{multline}
Since the additional factor of $r$ can be absorbed in $\nu$, Theorem~\ref{obs:allgood} follows from Eq.~(\ref{eq:eq3}).

Tight frames for which this is the case include those where the rank of the operators does not grow with $n$. The other extreme is given by the Pauli basis: From $\|w\|_2=1$ and $\|w\|=1/\sqrt{n}$ it follows that $\|w\|_1=\sqrt{n}$. Colloquially speaking, a small spectral norm implies a large trace norm and vice versa. Thus, we have two classes of tight frames (Fourier likes ones and the ones with small $1$-norm) for which efficient compressed sensing is efficiently possible. Because they represent in some sense the two extreme cases (flat spectra vs.\ concentrated spectra), we have some reason to believe that this is indeed true for \textit{any} tight frame.

\section{Certification}
\subsection{Ideal case}
Theorems \ref{obs:almostf} and \ref{obs:nf} show that efficient compressed sensing is possible in a vast number of situations. They are stated in the asymptotic regime for clarity but could be furnished with reasonable prefactors for finite Hilbert-space dimension $n$. 
However, when using compressed sensing in actual experiments, one encounters three main problems. 

\begin{itemize}
\item Firstly, the necessary number of measurements as calculated from the incoherence properties of the employed tight frame might still be too large to be feasible.

\item Secondly, repetition of the experiments to increase the probability of success to a satisfactory value may be expensive or difficult. 

\item Thirdly, it is unknown how close to low-rank the state actually is. After all, no assumptions
are made about the unknown input state.

\end{itemize}
The solutions to those problems is provided by certification. Instead of theoretically constructing some certificate based on $\rho$ with the help of the golfing scheme, we use the solution of the minimization problem $\sigma^*$ to explicitly check whether the conditions for Theorem \ref{lem:certified} are satisfied for $\sigma^*$. The candidate for the certificate can be calculated as
\begin{equation}
 \label{eq:certificate}
 Y=\R\PTp(\PTp\R\PTp)^{-1}{\rm\ sgn}\,\sigma^*
\end{equation}
 where $\PTp$ is obtained like $\PT$ but with $\rho$ replaced by $\sigma^*$ and $M^{-1}$ denotes the Moore-Penrose pseudo inverse of $M$. One can now check whether (\ref{eq:certified}) is fulfilled. If the conditions for Theorem \ref{lem:certified} are fulfilled and 
$\|\sigma^*\|_1=1$, the solution must be unique and equal to the state $\rho$, i.e. tomography was successful. 

\subsection{Errors and noise}
For compressed sensing to work in a realistic setting, the reconstruction procedure must be robust, i.e., small errors introduced by decoherence, errors stemming from imperfect measurements, and statistical noise due to the fact that every observable is only measured a finite number of times, should only lead to small errors in the reconstructed state. In addition, the Hilbert space might be infinite-dimensional. When the mean energy, and therefore, the mean photon number $N_{\rm mean}$, is finite, the error made by truncating the Hilbert space at photon number $N$ vanishes as 
\begin{equation}
\|\rho_{\rm trunc}-\rho\|_1\le3\sqrt{\frac{N_{\rm mean}}{N+1}}=\varepsilon
\end{equation}
which is shown in the appendix. This means that the expectation values with respect to the truncated state are close to the actually measured ones, i.e.,
\begin{equation}
	|\Tr(w\rho_{\rm trunc})-\Tr(w\rho)|\le\varepsilon,
\end{equation}	 
for all $w$ such that $\|w\|\leq 1$.

We assume that the observed data correspond to a matrix $\tilde{\rho}$ (not necessarily a state) with $\|\mathcal{P}_{\mathcal{R}}(\tilde{\rho}-\rho)\|_2\le\delta$ where $\rho$ is the low-rank, infinite-dimensional state, i.e., the errors made by truncating to a finite-dimensional Hilbert space are already included in $\delta$, and where we denote by $\mathcal{P}_{\mathcal{R}}$ the projection to the image of the sampling operator. Such a tube condition is satisfied with very high probability for realistic error models like Gaussian noise \cite{davidletter, noise}. We relax the conditions in (\ref{eq:reconstruction}) to 
\begin{equation}
	\|\mathcal{P}_{\mathcal{R}}(\sigma-\tilde{\rho})\|_2\le\delta. 
\end{equation}
The solution of the SDP might not be of low rank. Because a low-rank state is needed for the construction of the certificate $Y$, we truncate $\sigma^*$ to the $q$ largest eigenvalues (call this $\sigma^*_q$) and obtain $\mathcal{P}_{T'}$ as above. As $r={\rm rank}\,\rho$ is in general not known, one has to perform the truncation of $\sigma^*$ and the subsequent construction of the certificate $Y$ for $q=1,2,\ldots$ until a valid $Y$, as to be specified below, has been found. If this is not the case, the number of measurements was not enough and needs to be increased. 

To provide an error bound, we denote the $2$-norm error made by the truncation of $\sigma^*$ to rank $q$ by $\varepsilon$ and obtain from the triangle inequality
\begin{equation}
\|\mathcal{P}_{\mathcal{R}}(\sigma^*_q-\rho)\|_2=\|\mathcal{P}_{\mathcal{R}}(\sigma^*_q-\sigma^*)\|_2+\|\mathcal{P}_{\mathcal{R}}(\sigma^*-\tilde{\rho})\|_2+\|\mathcal{P}_{\mathcal{R}}(\tilde{\rho}-\rho)\|_2\le\varepsilon+2\delta.
\end{equation}
We calculate a candidate for a certificate as $Y=\R\PTp(\PTp\R\PTp)^{-1}{\rm\ sgn}\,\sigma^*_q$ where $T'$ is obtained from $\sigma^*_q$. If $Y$ is valid, i.e., $\|\mathcal{P}_{T'^\perp}Y\|\le 1/2$, and $\PTp\mathcal{P}_\mathcal{R}\PTp\ge(p/2)\PTp$ with $p=m/n^2$, then the proof of Theorem 7 in Ref.\ \cite{noise} yields the robustness result
\begin{equation}
\label{eq:robustness}
\|\sigma_q^*-\rho\|_2\le\left(4\sqrt{\frac{(2+p)n}{p}}+2\right)(2\delta+\varepsilon).
\end{equation}
By the equivalence of the norms, this also provides a $1$-norm bound at the expense of an additional factor $\sqrt{n}$. 

Thus, with no further assumption than $2$-norm closeness 
of the observations to the state of interest it is possible to obtain a certified reconstruction which is also close to the state of interest. In this sense, quantum compressed sensing can achieve
assumption-free certified quantum state reconstruction in the presence of errors.
This discussion applies to box errors, where each of the expectation values is assumed to
be contained in a certain interval. The discussion of other error models will be the subject
of forthcoming work.

\section{Universal quantum compressed sensing}

\subsection{Universal quantum state reconstruction}

The preceding discussion has focused on claims of the following form: 
\begin{quotation}\noindent
For every low-rank state $\rho$, most choices of the observables $w_1,\ldots,w_m$ can be used to successfully reconstruct $\rho$. 
\end{quotation}
In some situations, however, one can actually prove a much stronger statement, in which the order of the quantifiers is reversed: 
\begin{quotation}\noindent
Most choices of the observables $w_1,\ldots,w_m$ will have the property that, for every low-rank state $\rho$, the observables $w_1,\ldots,w_m$ can be used to successfully reconstruct $\rho$. 
\end{quotation}
This is known as \textit{universal} reconstruction; more simply, it says that a fixed set of observables $w_1,\ldots,w_m$ can distinguish among all low-rank states \textit{simultaneously}. Besides being of conceptual interest, universal reconstruction also implies stronger error bounds for reconstruction from noisy data.

Formally, we say that our method performs universal compressed sensing if $p_{fu} < 1/2$, where $p_{fu}$ is the ``universal'' failure probability. That is, we define $p_{fu}$ to be the probability (with respect to the choice of observables $w_1,\ldots,w_m$) that there exists a state $\rho$ (with dimension $n$ and rank $r$) such that the method fails with probability $> 1/2$ (where this last probability is taken with respect to the random measurement outcomes). 

\begin{definition}[Efficient universal quantum compressed sensing]
\label{def:efficientuniversal}
Universal compressed sensing (with dimension $n$ and rank $r$) is regarded as 
efficient if: the number of measured observables satisfies $m=O(nr\,\polylog(n))$, and the probability of failure satisfies $p_{fu} < 1/2$. 
\end{definition}

\subsection{Universal reconstruction using any Fourier-type tight frame}

In this section, we show that measurements using a Fourier-type tight frame lead to efficient universal quantum compressed sensing. This result can be viewed as a companion to Theorem \ref{obs:almostf}. Essentially, it says that, by using a slightly larger number of measurements (by a polylog($n$) factor), one can construct (with high probability) a single, \textit{fixed} set of measurements that can reconstruct \textit{all} possible states of rank $r$ and dimension $n$. In addition, universal reconstruction implies very strong error bounds, in the case of noisy data; we will discuss this in the following section.

\begin{theorem}[Universal reconstruction]
\label{obs:universal}
Let $(w_\alpha)_{\alpha \in S}$ be a tight frame. 
Let $\nu=O({\rm polylog}(n))$, and suppose that, for all $\alpha \in S$, $\|w_\alpha\|^2 \leq \nu/n$.
Then efficient universal compressed sensing (for states of rank $r$ and dimension $n$) is possible. 
\end{theorem}

This proof of this theorem is a straightforward generalization of \cite{yikai}. First, we define the sampling operator to be $\calA:\: \CC^{n\times n} \rightarrow \RR^m$, 
\begin{equation}
(\calA(\sigma))_i = \frac{n}{\sqrt{m}} (w'_i, \sigma), \quad i = 1,\ldots,m.
\end{equation}
This is related to the notation used in previous sections by $\calA^\dagger \calA = \R$. (As before, the observables $w'_1,\ldots,w'_m$ are sampled independently from the distribution $\mu$ on the tight frame, and $(A,B) = \Tr(A^\dagger B)$ is the Hilbert-Schmidt inner product.)

A key tool in the proof is the \textit{restricted isometry property} (RIP) \cite{RFP}.  We say that $\calA$ satisfies the RIP if there exists some constant $\delta \in [0,1)$ such that, for all rank-$r$ $n$-dimensional states $\sigma$, 
\begin{equation}\label{eq:rip}
(1-\delta) \|\sigma\|_2 \leq \|\calA(\sigma)\|_2 \leq (1+\delta) \|\sigma\|_2.
\end{equation}
In geometric terms, the set of all low-rank states forms an $O(rn)$-dimensional manifold in $\CC^{n\times n}$, and $\calA$ satisfies the RIP if it embeds this manifold into $\CC^m$, with constant-factor distortion.

The importance of the RIP stems from the following fact:  when $\calA$ satisfies RIP, one can reconstruct any low-rank state $\rho$ from noiseless data $\calA(\rho)$, by solving a trace-minimization convex program:
\begin{equation}
\min \|\sigma\|_1\,,\quad{\rm subject\,\,to}\,\,\mathcal{A}(\sigma)=\mathcal{A}(\rho).
\label{eq:reconstruction2}
\end{equation}
This follows from a standard argument of \cite{RFP}. This result can be generalized to the case of noisy data; we will discuss this in the following section.

It now remains to prove that, when the observables $w_\alpha$ are chosen from a Fourier-type tight frame (i.e., they satisfy $\|w_\alpha\|^2 \leq \nu/n$), the sampling operator $\calA$ satisfies RIP with high probability.  Intuitively, one first shows that, for any particular low-rank state $\sigma$, and a random choice of measurements $w'_1,\ldots,w'_m$, the sampling operator $\calA$ satisfies equation (\ref{eq:rip}) with high probability. After this comes the main part of the argument. Let $p_f(\sigma)$ denote the probability of failure on a given state $\sigma$. One now needs to upper-bound the probability of a failure on any one of the states $\sigma$. The simplest approach is to assume that the failure events are disjoint, and so the probabilities sum up --- this is the union bound, and it does not give a useful bound in this case. Instead, one uses an ``entropy argument'' that exploits the fact that failure events are not disjoint: failures on nearby states are correlated. 

Formally, the entropy argument is carried out using Gaussian processes and Dudley's inequality (following \cite{rudelson-vershynin, candes-tao-universal}), and using bounds on covering numbers of the trace-norm ball due to \cite{guedon-et-al}. The proof is essentially the same as in \cite{yikai}; the original proof in \cite{yikai} handles the case where the $w_\alpha$ form an incoherent orthonormal basis, but the same proof goes through unchanged for a Fourier-type tight frame. This shows that, if the number of measurements satisfies $m \geq C\nu rn\log^6 n$ (for some constant $C$), then with high probability the sampling operator $\calA$ satisfies the RIP (for rank $r$ and dimension $n$). 

\subsection{Robust reconstruction from noisy data}

More interestingly, RIP implies strong error bounds in the case of noisy data \cite{cpTightOracleIneq}. We sketch the basic idea here. Suppose one observes $y = \calA(\rho) + z$, where $z$ denotes a noise component. Then one can replace (\ref{eq:reconstruction2}) with other estimators, such as the matrix Dantzig selector: 
\begin{equation}
\min \|\sigma\|_1 \quad \text{such that} \quad \| \calA^\dagger (y-\calA(\sigma)) \| \leq \lambda,
\end{equation}
or the matrix Lasso:
\begin{equation}
\min \tfrac{1}{2} \| \calA(\sigma)-y \|_2^2 + \mu \|\sigma\|_1.
\end{equation}
(See Ref.\ \cite{cpTightOracleIneq} for details about setting the parameters $\lambda$ and $\mu$.) 

When the noise vector $z$ is normally distributed, one can show particularly nice error bounds. These hold even for states $\rho$ that are \textit{full-rank} (though $\rho$ must at least have decaying eigenvalues, for the bounds to be useful) \cite{cpTightOracleIneq} (see also \cite{yikai}). Suppose that $\rho$ is arbitrary, and one simply assigns some value for $r$, and measures $m = O(\nu rn \log^6 n)$ observables. Then let $\sigma^*$ denote the solution returned by either of the above estimators. Intuitively, one expects that $\sigma^*$ should reconstruct the first $r$ eigenvectors of $\rho$. One can prove a bound that is consistent with this intuition: the squared $2$-norm error $\| \sigma^* - \rho \|_2^2$ will be nearly proportional (up to log factors) to the total variance of the noise acting on the first $r$ eigenvectors of $\rho$, plus the squared $2$-norm of the ``tail'' of $\rho$ (consisting of its last $n-r$ eigenvectors).

\section{Applications}

We now demonstrate how our theory can be applied to some common experimental setups in quantum optics. We show how pointwise measurements of the Wigner function, and histograms obtained using homodyne detection, can be expressed as measurements using tight frames, and generalized tight frames. Furthermore, we propose efficient compressed sensing schemes (with Fourier-type tight frames) using these measurements. 

\subsection{Homodyne detection}
\label{sus:homodyne}

The most common way to do quantum state tomography on continuous-variable light modes is based on homodyne detection, which is done by combining the light field with a mode in a strong coherent state, called the local oscillator, in an interferometer and measuring the difference of the intensities on the two output ports \cite{leon,leonletter,us}. 
This amounts to sampling $x \in \RR$ according to the one-dimensional probability distribution given by the Radon transform (at angle $\theta$) of the Wigner function, i.e.,
\begin{equation}
\label{eq:radon}
\PR_\theta(x)=\int W(x\cos\theta-p\sin\theta,x\sin\theta+p\cos\theta){\rm d}p.
\end{equation}
The angle $\theta$ is chosen by phase-shifting the mode with respect to the local oscillator. 

For a 
general quantum state with maximal photon number $N$, $N+1$ equidistant choices of $\theta\in[0,\pi)$ are sufficient and necessary to reconstruct the state by an inverse Radon transform
of Eq.\ (\ref{eq:radon}) or by using pattern functions \cite{leon,leonletter}. 
Here we show how these measurements can be described by a generalized tight frame. A tight frame by itself does not suffice because here every measurement setting, i.e., every choice of $\theta$, does not only give a single number as a result but an entire 
distribution $\PR_\theta$. 

A key observation is that the Fourier transform of the probability distribution (\ref{eq:radon}) is identical to the characteristic function, i.e., the Fourier transform of the Wigner function, written in radial coordinates. We define
\begin{equation}
\tilde{W}(u,v)=\int\md x\,\md p\,W(x,p)\exp[-i(ux+vp)]
\label{eq:Wtilde}
\end{equation}
which fulfills 
\begin{equation}
	\tilde{\Pr}_\theta(\zeta)=\tilde{W}(\zeta\cos\theta,\zeta\sin\theta) 
\label{eq:wigner-ft-polar}
\end{equation}
where $\tilde{\Pr}_\theta(\zeta)=\int dx\Pr_\theta(\zeta)\exp(-i\zeta x)$. 

This allows us to write the projector (corresponding to measurement setting $\theta$ and outcome $\zeta$) as
\begin{equation}
	\left(\mathcal{P}_\theta(\zeta)\right)_{(i,j),(k,l)}=\tilde{W}_{|j\rangle\langle i|}(\zeta\cos\theta,\zeta\sin\theta)
	\tilde{W}^*_{|l\rangle\langle k|}(\zeta\cos\theta,\zeta\sin\theta)\,.
\end{equation}
Because choosing a measurement setting does not mean choosing values for $\theta$ and $\zeta$, but rather only choosing a phase $\theta$ and obtaining a whole ``slice'' of the characteristic function, the operator corresponding to a measurement setting is 
\begin{equation}
\label{eq:Ptheta}
\mathcal{P}_\theta=\int\md\zeta\,\mathcal{P}_\theta(\zeta)\,.
\end{equation}
It is easy to check that $\mathcal{P}_\theta$ fulfills
\begin{equation}
\frac{1}{\pi}\int_0^\pi{\rm d}\theta\,\mathcal{P}_\theta=\frac{\id}{n^2},
\end{equation}
which implies that it satisfies Definition~\ref{def:tfcondition2} and forms a generalized tight frame.

\subsection{Efficient compressed sensing using homodyne measurements}
\label{sus:efficienthomodyne}

In the previous subsection, we have introduced the generalized tight frame corresponding to homodyne detection. This can be combined with the convex program in equation (\ref{eq:reconstruction}) to perform state reconstruction. In Section \ref{sec:numerics}, we show by means of a numerical simulation that this procedure performs well in practice. However, our theoretical analysis does not apply to this procedure, due to the generalized tight frame; it would be interesting to try to extend our theoretical results to this case.

In this section, we present a different way of using homodyne detection to reconstruct low-rank states, which is a little less direct, but does have a rigorous guarantee of success. We will do three things. First, we will show how homodyne measurements can be used to estimate expectation values of displacement operators. Then, we will use (scaled) displacement operators to construct a tight frame. Finally, we will show that this tight frame has Fourier-type incoherence. By combining these pieces, we will then get an efficient compressed sensing scheme. 

Before continuing, we note that $D(\alpha)$ cannot be directly measured as it is not Hermitian. However, one can also use 8-port homodyning to directly measure the observables $|\alpha\rangle\langle\alpha|=D(\alpha)|0\rangle\langle 0|D^\dag(\alpha)$ \cite{schleich}. Because the experimental effort is higher, compared to standard homodyning, we will not discuss this scheme here.

Define the displacement operators 
\begin{equation}
\label{eq:disop}
D(\alpha) = e^{-|\alpha|^2/2} e^{\alpha a^\dg} e^{-\alpha^* a}, \qquad \alpha\in\CC.
\end{equation}
Note that we have the identities $D(\alpha) = e^{\alpha a^\dg - \alpha^* a} = e^{|\alpha|^2/2} e^{-\alpha^* a} e^{\alpha a^\dg}$.

Now recall the definition of the characteristic function \cite{scully-zubairy}:
\begin{equation}
C^{(s)}(\beta) = \Tr(e^{i\beta a^\dg + i\beta^* a} \rho), \qquad \beta\in\CC.
\end{equation}
Setting $\alpha = i\beta$, we see that $C^{(s)}(\beta)$ is precisely the expectation value of the displacement operator $D(\alpha)$. On the other hand, $C^{(s)}(\beta)$ is also equal to $\tilde{W}(\beta)$, the (two-dimensional) Fourier transform of the Wigner function $W(\xi)$. This in turn is related, via equation (\ref{eq:wigner-ft-polar}), to the probability distribution $\PR_\theta(x)$, which we can sample using homodyne detection.

Thus, we can estimate the expectation value of a displacement operator $D(\alpha)$ as follows: set $\beta = -i\alpha$, and make homodyne measurements with phase angle $\theta = \arg(\beta)$. This produces several points $x_1,\ldots,x_\ell \in \RR$ sampled from the distribution $\PR_\theta(x)$. Then set $\zeta = |\beta|$, and compute $\frac{1}{\ell} \sum_{i=1}^\ell \exp(-i\zeta x_i)$. This gives an estimate for $\tilde{\PR}_\theta(\zeta) = \tilde{W}(\beta) = C^{(s)}(\beta)$, which is the desired expectation value.

Note that a lossy detector (i.e., one with efficiency less than 1) has the effect of convolving the true Wigner function $W(\xi)$ with a Gaussian, to produce the empirically observed Wigner function \cite{raymer-beck}. This is equivalent to pointwise multiplying the characteristic function $C^{(s)}(\beta)$ with a Gaussian envelope. We can compensate for this by re-scaling $C^{(s)}(\beta)$ at each point $\beta$, provided that our raw estimates of $C^{(s)}(\beta)$ are sufficiently precise, and the detector efficiency is not too poor.

Next, we will construct a tight frame using the displacement operators $D(\alpha)$. Note that the $D(\alpha)$ form an orthonormal basis for the state space \cite{scully-zubairy}:
\begin{equation}\label{eqn-cmplt}
\rho = \frac{1}{\pi} \int_{\CC} D(\alpha) \Tr(D(\alpha)^\dg \rho){\rm d}\alpha, \qquad \text{for all states $\rho$},
\end{equation}
where we are taking a $2$-dimensional integral over the complex plane. Now suppose we sample $\alpha$ from a $2$-dimensional Gaussian distribution on the complex plane with width $\sigma$ (which we will choose later). This distribution has probability density 
\begin{equation}
\label{eq:PG}
P_G(\alpha) = \tfrac{1}{2\pi\sigma^2} e^{-|\alpha|^2/2\sigma^2}.
\end{equation}
Define scaled displacement operators 
\begin{equation}
\tilde{D}(\alpha) = \sqrt{2}\sigma e^{|\alpha|^2/4\sigma^2} D(\alpha).
\end{equation}
Then we can rewrite (\ref{eqn-cmplt}) as 
\begin{equation}
\rho = \int_{\CC} \tilde{D}(\alpha) \Tr(\tilde{D}(\alpha)^\dg \rho) P_G(\alpha){\rm d}\alpha, \qquad \text{for all states $\rho$}.
\end{equation}
This is (up to normalization) a tight frame for the full, infinite-dimensional state space.

In fact, we are only interested in the finite-dimensional subspace consisting of states with at most $N$ photons; this subspace is isomorphic to $\CC^{(N+1)\times(N+1)}$. So we will truncate the above operators. Let $\Pi_N$ be the projector onto $\text{span}\set{\ket{0},\ket{1},\ldots,\ket{N}}$ (where the $\ket{j}$ are Fock basis states). Then define truncated displacement operators 
\begin{equation}
D_N(\alpha) = \Pi_N D(\alpha) \Pi_N, \qquad \text{and} \qquad
\tilde{D}_N(\alpha) = \Pi_N \tilde{D}(\alpha) \Pi_N.
\end{equation}
Then the operators $w_\alpha = \tfrac{1}{N+1} \tilde{D}_N(\alpha)$ form a tight frame for $\CC^{(N+1)\times(N+1)}$, as desired:
\begin{equation}
\tfrac{1}{(N+1)^2} \rho = \int_{\CC} w_\alpha \Tr(w_\alpha^\dg \rho) P_G(\alpha) d\alpha, \qquad \text{for all $\rho \in \CC^{(N+1)\times(N+1)}$}.
\end{equation}

Finally, we set $\sigma = \sqrt{2N\log(1+4N)}$, and we claim that the above tight frame $\set{w_\alpha}$ is Fourier-type incoherent, in the sense of Theorems \ref{obs:almostf} and \ref{obs:universal}. More precisely, we claim that 
\begin{equation}\label{eqn-onb}
\norm{\tilde{D}_N(\alpha)} \leq \sqrt{2}e\sigma = 2e\sqrt{N\log(1+4N)}, \qquad \text{for all $\alpha\in\CC$};
\end{equation}
we will prove this below. This directly implies 
\begin{equation}
\norm{w_\alpha} \leq \frac{2e\sqrt{\log(1+4N)}}{\sqrt{N}}, \qquad \text{for all $\alpha\in\CC$}.
\end{equation}
Then, by Theorems \ref{obs:almostf} and \ref{obs:universal}, we have an efficient compressed sensing scheme.

We now show why (\ref{eqn-onb}) holds. First, note that while the displacement operators $D(\alpha)$ are unitary, the scaled operators $\tilde{D}(\alpha)$ are unbounded. However, when $\alpha$ is small, this is not a problem. In particular, when $|\alpha| \leq 2\sigma$, we can just use the trivial bound 
\begin{equation}
\norm{\tilde{D}_N(\alpha)} \leq \norm{\tilde{D}(\alpha)} \leq \sqrt{2}\sigma e^{|\alpha|^2/4\sigma^2},
\end{equation}
which implies (\ref{eqn-onb}).

It remains to consider the case where $|\alpha| > 2\sigma$. In this case, $\tilde{D}(\alpha)$ is large, but it acts mostly on states with more than $n$ photons, so the truncated operator $\tilde{D}_N(\alpha)$ is small. Using a straightforward calculation, we can bound $\tilde{D}_N(\alpha)$ in the $2$-norm, which implies (\ref{eqn-onb}). See the appendix for details.

\subsection{Pointwise measurements of the Wigner function}

A quantum state $\rho$ of a single optical mode can be represented in phase space by a real 
Wigner function $W_\rho:\rr^2\rightarrow \rr$ \cite{schleich}. For a single mode it is given by, c.f.\ Ref.~\cite{vogel,vogel2},
\begin{equation}
\label{eq:pointwisewigdef}
W_\rho(\xi)=\frac{2}{\pi}\,\Tr\left((-1)^{\hat{n}}\,D(\xi)^\dag\rho D(\xi)\right)
\end{equation}
where $(-1)^{\hat{n}}$ is the parity operator where $\xi=(x,p)\in\mathbbm{R}^2$, $D(\xi)$ is the displacement operator which becomes the one defined in (\ref{eq:disop}) by setting $\alpha=(1/\sqrt{2})(\xi_1+i\xi_2)$. With the same convention, we will, whenever it is convenient, regard the Wigner function as a function of a complex variable. 

Eq.~(\ref{eq:pointwisewigdef}) allows pointwise measurement of the Wigner function by a displacement in phase space followed by a measurement of the parity of the photon number. This has already been experimentally performed for optical fields in a cavity \cite{haroche} and for pulsed single photons (for the special case of a rotationally invariant state) \cite{wignerpointwise}. 
We consider a single mode containing up to $N$ photons and, therefore, Hilbert space dimension $n=N+1$. Measuring the Wigner function at a point $\alpha$ amounts to a measurement of the observable
\begin{equation}
\label{eq:defwxi}
w_\alpha =\sqrt{2\pi}\frac{2}{\pi}D(\alpha)(-1)^{\hat{n}}D^\dag(\alpha).
\end{equation}
We make again use of the probability density $P_G$ of Eq.\ (\ref{eq:PG}) and define scaled, truncated operators $\tilde{w}_\alpha=n^{-1}P_G(\alpha)^{-1/2}\Pi_Nw_\alpha\Pi_N$. They form a tight-frame on the truncated Hilbert space when the sampling is performed according to $P_G$. 

We now proceed exactly as in the previous section to show that the operator norm of the $\tilde{w}$ is small enough for the Fourier type incoherence property of Theorem \ref{obs:almostf}. We do not give explicit constants but focus on the asymptotic scaling in $n$. If $|\alpha|\le2\sigma$, we get the bound $\|\tilde{w}_\alpha\|\le 4e\sigma/n$. We will show that if we set $\sigma=\sqrt{n}\log n$ one has $\|\exp(|\alpha|^2/(2\sigma^2))w_\alpha\|\le 1$ for all $\alpha$ with $|\alpha|>2\sigma$ whenever $n$ is large enough which implies the requirements of Theorem \ref{obs:almostf}. We need the matrix elements $\langle l|w_\alpha|k\rangle=W_{|k\rangle\langle l|}(\alpha)$. To calculate them, first let $\xi=(x,p)$ and recall the definition of the Wigner function \cite{schleich}:
\begin{equation}
	W_{|l\rangle\langle k|}(x,p)=\frac{1}{\pi}\int\md y\,\psi_l^*(x+y)\psi_k(x-y)e^{2ipy} .
\end{equation}
where we remember the identification $\alpha=(1/\sqrt{2})(x+ip)$. Inserting the eigenfunctions of the harmonic oscillator $\psi_i$, using the properties of the 
occurring Hermite polynomials, and performing the integral allows to write
\begin{equation}
\label{eq:wigpoint}
	W_{|l\rangle\langle k|}(x,p)=\frac{(-1)^{l+k}e^{x^2}}{\pi\sqrt{2^{l+k}l!k!}}\frac{\partial^{l+k}}{\partial x^k\partial x'^l}G(x,x,p')\Biggr|_{x'=x}
\end{equation}
with the generating function
\begin{equation}
	\label{eq:genfun}
	G(x,x',p)=e^{-p^2+2ip(x-x')-2xx'}.
\end{equation}
From this, one gets the bound, which is by no means tight but strong enough, $|W_{|k\rangle\langle l|}(\alpha)|\le n^n(2|\alpha|)^n\exp(-2|\alpha|^2)$ which allows us to write
\begin{align}
\|\exp(|\alpha|^2/(2\sigma^2))w_\alpha\|\le&\|\exp(|\alpha|^2/(2\sigma^2))w_\alpha\|_2\nonumber\\
\le&\exp\left(-2|\alpha|^2+(n+2)\log n+2n\log(2|\alpha|)+\frac{|\alpha|^2}{2\sigma^2}\right).
\label{eq:wignerbound}
\end{align}
We now set $\sigma=\sqrt{n}\log n$ and get, for large enough $n$, a bound valid for all $\alpha$ with $|\alpha|>2\sigma$ which reads
\begin{equation}
\label{eq:wignerbound2}
\|\exp(|\alpha|^2/(2\sigma^2))w_\alpha\|\le\exp\left(-2n\log^2n+(n+2)\log n+2n\log(4\sqrt{n}\log n)\right).
\end{equation}
As the first term in the exponent grows fastest, one has $\|\exp(|\alpha|^2/(2\sigma^2))w_\alpha\|\le 1$ for sufficiently large $n$. Thus, there is some $C>0$ such that $\|\tilde{w}_\alpha\|\le C\log n/\sqrt{n}$ which means that the pointwise Wigner function measurement is of Fourier type and, therefore, can be used for efficient compressed sensing.

\section{Numerical examples}
\label{sec:numerics}
We now present some examples which show the performance of certified compressed sensing for randomly chosen states. We demonstrate the method for small-dimensional noiseless states and defer a detailed analysis of the method, especially in the presence of noise and decoherence, to a subsequent publication. For small systems, the condition $c_3<1$ in Theorem~\ref{lem:certified} is hard to satisfy. However, the conditions for uniqueness can be replaced by (a') $c_1 := \|\PT Y-{\rm sgn}\,\rho\|_2=0$ and (b') $c_2 := \|\PT^\perp Y\|<1$, discarding the condition on $c_3$, because these conditions imply that the expression in (\ref{eq:pinching2}) is positive, which guarantees that any feasible change in the solution will be $1$-norm increasing. 

\begin{figure}
\unitlength1mm
\begin{center}
\begin{picture}(70,53)
\put(0,0){\includegraphics[width=7cm]{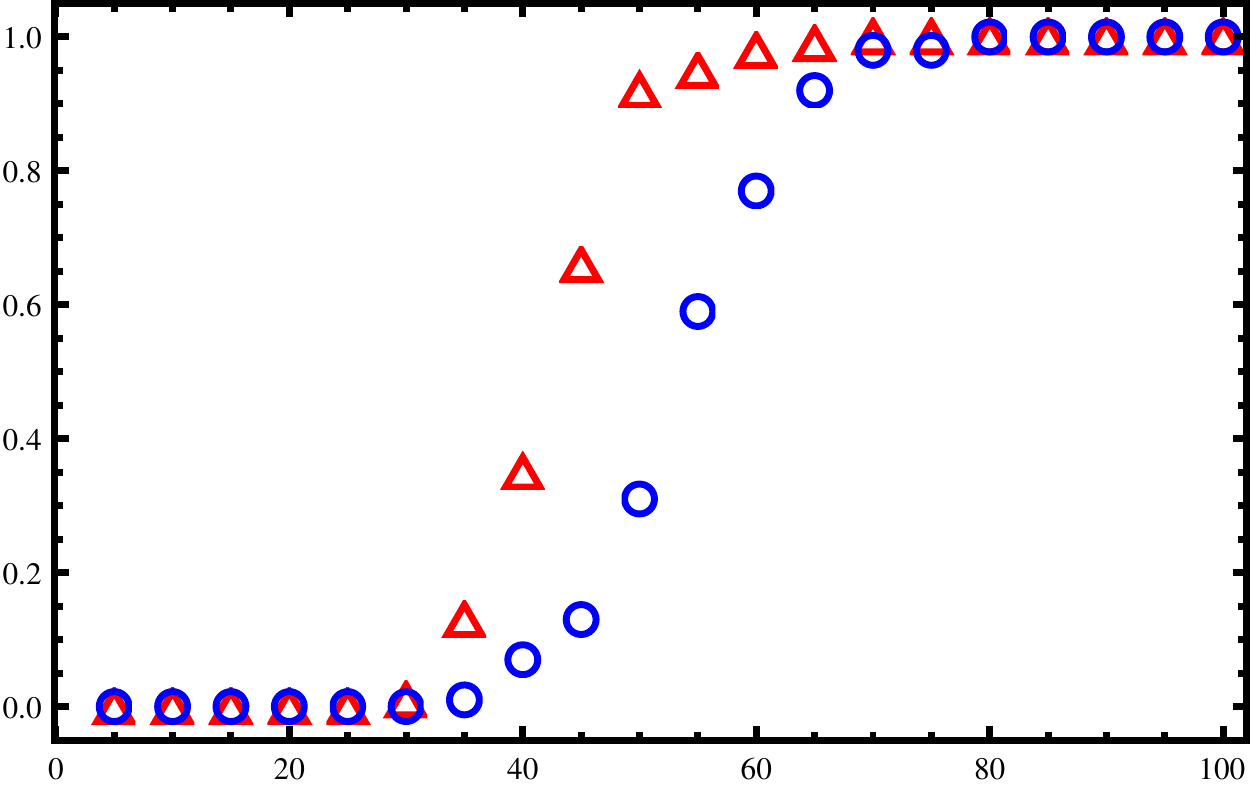}}
\put(35,-3){\small $m$}
\put(0,45){\small $P$}
\end{picture}
\end{center}
\caption{\label{fig:Pauli} (color online) Reconstruction of a random pure state on $4$ qubits by Pauli-measurements. Red triangles: Probability of successful state recovery. Blue circles: Probability of successful certification. }
\end{figure}

Figure~\ref{fig:Pauli} demonstrates certified compressed sensing for the very important case of the Pauli basis. It is clearly visible that the certificate is only a sufficient condition and not a necessary one as it is possible that the reconstruction is successful but no valid certificate is produced. It is also apparent that the overhead in the number of queries needed for certification is actually quite reasonable. 

\begin{figure}
\unitlength1mm
\begin{center}
\begin{picture}(70,53)
\put(0,0){\includegraphics[width=7cm]{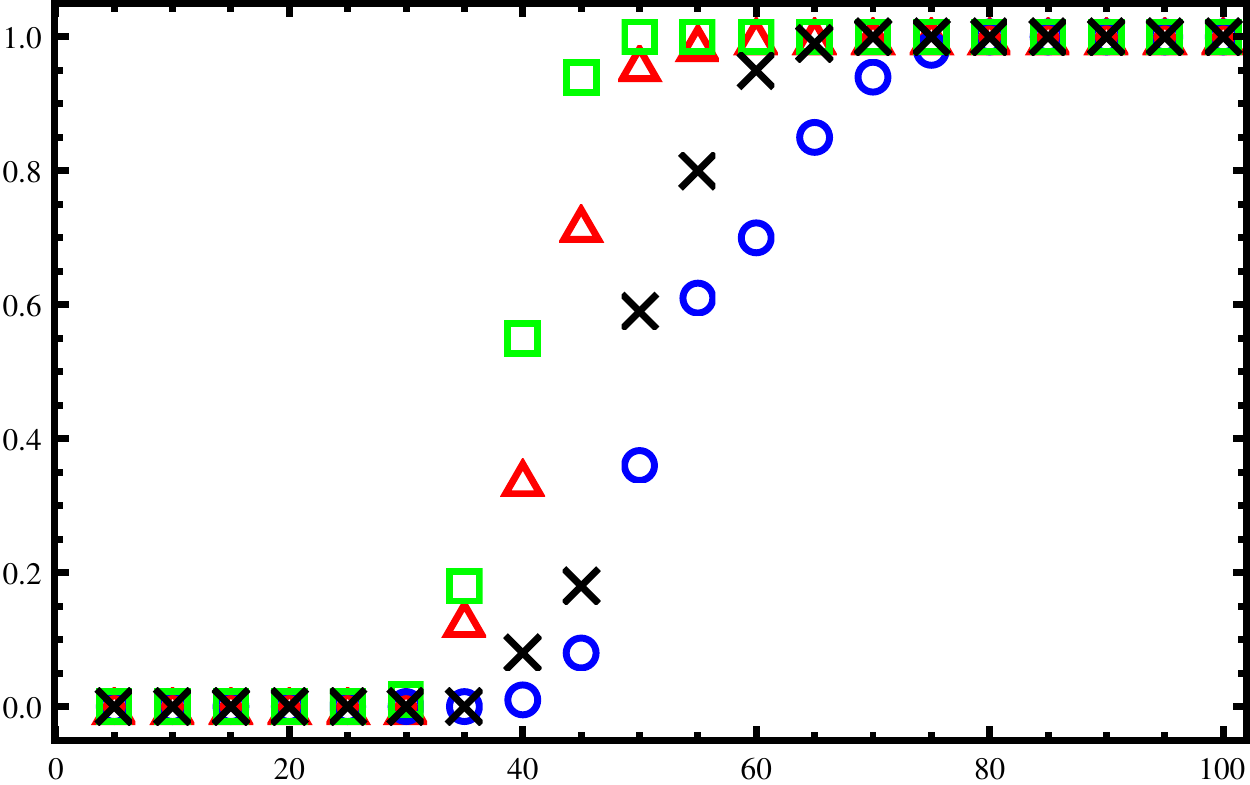}}
\put(35,-3){\small $m$}
\put(0,45){\small $P$}
\end{picture}
\end{center}
\caption{\label{fig:Example} (color online) Reconstruction of a pure state on $4$ qubits. Red triangles (blue circles): Probability of successful state recovery (certification) for \textit{local} random measurements. Green squares (black crosses): Successful state recovery (certification) for \textit{global} random measurements.}
\end{figure}

For the tight frame consisting of all Hermitian matrices, as shown in Figure~\ref{fig:Example}, it is interesting to note that taking global random observables performs superior to taking tensor products of local random observables. The intuitive reason for this is provided by concentration of measure. By considering a distribution of observables which is invariant under the action of the unitary group on the full system, the proportion of observables that are not Fourier-like, i.e., whose operator norms are too large, is much smaller. Thus, more information is obtained per observable which leads to a faster reconstruction. 

\begin{figure}
\unitlength1mm
\begin{center}
\begin{picture}(70,53)
\put(0,0){\includegraphics[width=7cm]{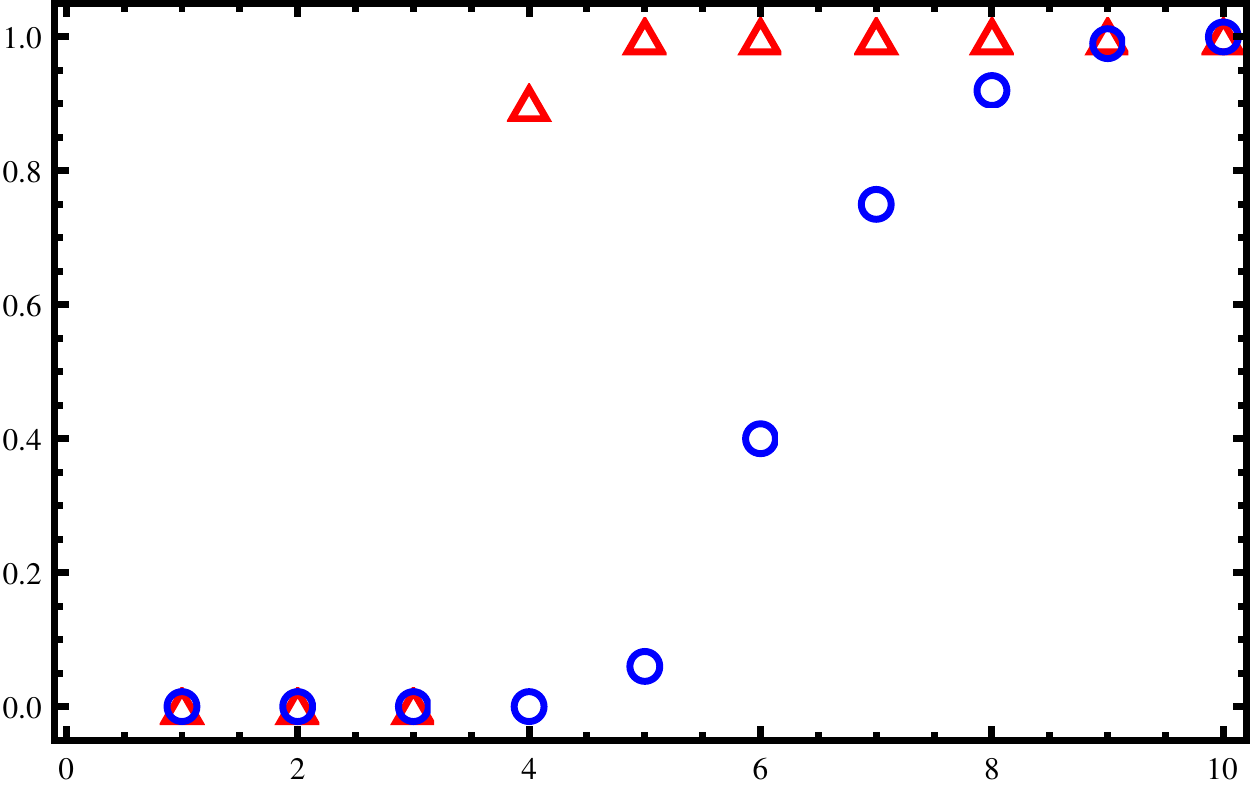}}
\put(35,-3){\small $m$}
\put(0,45){\small $P$}
\end{picture}
\end{center}
\caption{\label{fig:Hom} (color online) Reconstruction of a random state with rank $5$ on $3$ modes with up to $2$ photons each by optical Homodyne detection. Red triangles: Probability of successful state recovery. Blue circles: Probability of successful certification. }
\end{figure}

\begin{figure}
\unitlength1mm
\begin{center}
\begin{picture}(70,53)
\put(0,0){\includegraphics[width=7cm]{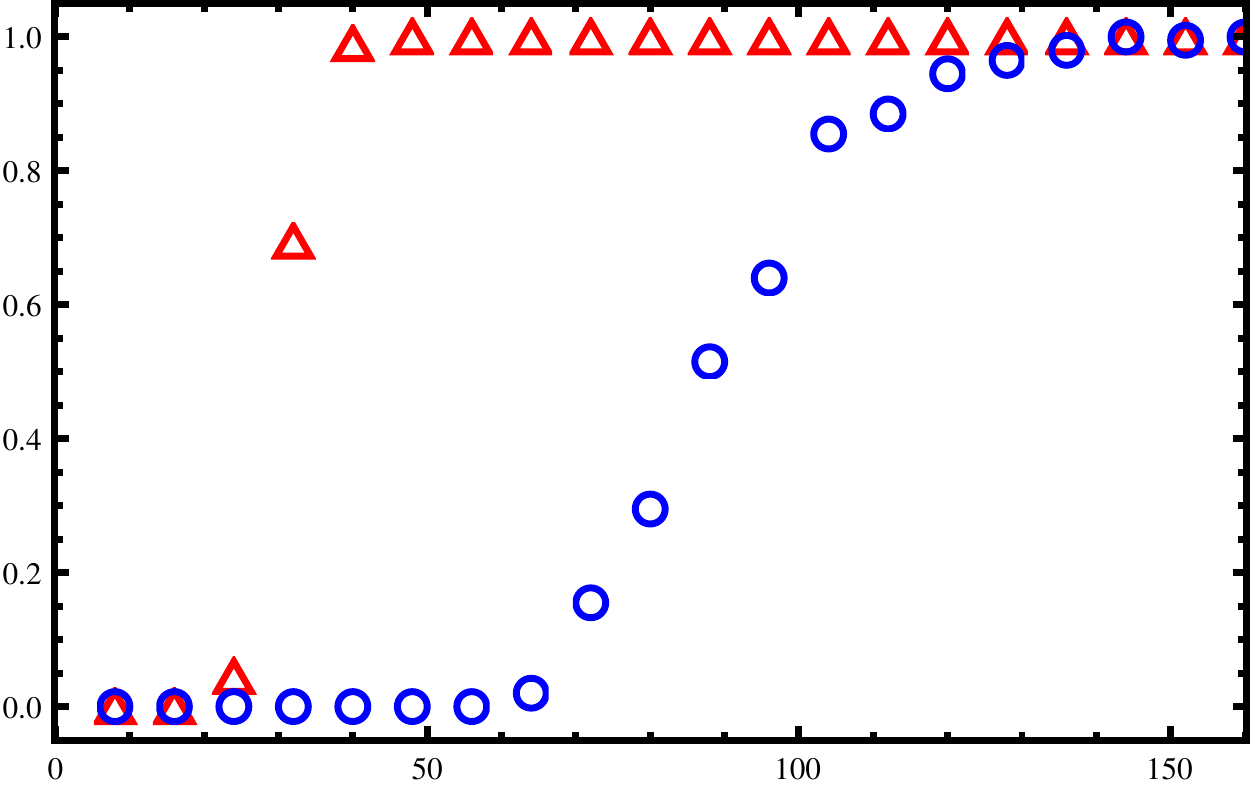}}
\put(35,-3){\small $m$}
\put(0,45){\small $P$}
\end{picture}
\end{center}
\caption{\label{fig:displacementops} (color online) Reconstruction of a random state of a single optical mode, truncated at the $15$-th number state, by measuring expectation values of $2$-norm normalized displacement operators $D(\alpha)$ where $|\alpha\|$ is chosen uniformly at random between $0$ and $5$ while ${\rm arg}(\alpha)$ is chosen uniformly between $0$ and $2\pi$. Red triangles: Probability of successful state recovery. Blue circles: Probability of successful certification.} 
\end{figure}

Figure~\ref{fig:Hom} illustrates that compressed sensing also works using optical homodyne detection with a generalized tight frame, c.f.\ Subsection~\ref{sus:homodyne}. In Figure~\ref{fig:displacementops}, we show the reconstruction of a single mode optical state based on the measurement of expectation values of displacement operators as discussed in Subsection~\ref{sus:efficienthomodyne}.

\section{Summary}
In this article, we have presented a general theory of quantum state tomography for continuous-variable systems using compressed sensing. We have used tight frames to describe continuous measurement families, which are very natural in a plethora
of physical situations. We have shown how our theory applies to prominent and frequently
used techniques in quantum optics, in particular, pointwise measurements of the Wigner function, and homodyne detection. 
\begin{itemize}
\item We have explored different incoherence properties sufficient for efficient compressed sensing. Improved results using Fourier-type tight frames were presented in Theorem \ref{obs:almostf}. Also, it was shown in Theorem \ref{obs:nf} that for every tight frame whose operators fulfill 
\begin{equation}
\|w_\alpha\|_1=O({\rm polylog}(n)), 
\end{equation}
most states (i.e., all but a proportion $1/{\rm poly}(n)$ thereof) can be reconstructed from merely $O(n\,{\rm polylog}(n))$ expectation values. It would be interesting to extend these results to generalized tight frames.

\item We have introduced the idea of certified compressed sensing which allows to get rid of all assumptions and guarantee successful state reconstruction a posteriori. This assumption-free certified quantum state reconstruction is possible even in the presence of errors. 

\item Furthermore, we have shown universal compressed sensing results for any Fourier-type tight frame in Theorem \ref{obs:universal}. This implies strong error bounds in the case of noisy data.

\item We have presented numerical results showing the practical (non-asymptotic) performance of these methods. 
It would be interesting to investigate this further, in particular to other types of feasible measurements,
and to apply these ideas on other physical systems as well.

\end{itemize}

\section{Acknowledgements}
We would like to thank Jukka Kiukas for comments and Earl Campbell for discussions. 
YKL thanks Scott Glancy and Manny Knill for many helpful explanations and suggestions.
This work was supported by the EU (Qessence, Minos, Compas), the BMBF (QuOReP), and the EURYI.
Contributions to this work by NIST, an agency of the US government, are not subject to copyright laws.

\section*{Appendix}

\subsection*{Properties of the $\chi_k^2$-distribution}
In order to be self-contained, we repeat two simple bounds to the tails of a $\chi_k^2$ distributed 
random variable $X$ which can be found in Ref.\ \cite{chisquared}. A right-sided bound is
\begin{equation}
\label{eq:chi21}
\PP\left(X-k>2\sqrt{kx}+2x\right)\le e^{-x},
\end{equation}
while a left-sided one is
\begin{equation}
\label{eq:chi22}
\PP\left(k-X>2\sqrt{kx}\right)\le e^{-x}\,.
\end{equation}

\subsection*{Random vectors on a sphere}

A random vector $v\in\mathbb{C}^n$ on a sphere can be obtained by choosing an vector $\bar{v}\in\mathbb{R}^{2n}$ with Gaussian entries and normalizing. Doing so yields
\begin{equation}
	\PP\left(|v_i|\ge\frac{\delta}{\sqrt{n}}\right)\le\PP\left(|\bar{v}_i|>\frac{\delta\varepsilon}{\sqrt{n}}\right)+\PP\left(\|\bar{v}\|<\varepsilon\right)\,.
\end{equation}
To bound the first term, one can use (\ref{eq:chi21}), obtaining
\begin{equation}
	\PP\left(|\bar{v}_i|>\frac{1}{\varepsilon\sqrt{n}}\right)\le \exp\left(-\frac{\delta^2\varepsilon}{2}\right)
\end{equation}
while for the second terms the inequality (\ref{eq:chi22}) leads to
\begin{equation}
	\PP\left(\|\bar{v}\|^2<1-y\right)<\exp\left(-\frac{y^2n}{2}\right).
\end{equation}
Setting $\varepsilon=1/2$ finally gives
\begin{equation}
	\PP\left(|v_i|>\delta/\sqrt{n}\right)\le 2\exp\left(-\frac{\delta^2}{8}\right)\,.
\end{equation}

\subsection*{Proof of Lemma \ref{lem:probtheory}}

Proof: From 
\begin{equation}\label{eq:p}
\mathbbm{P}\left(\mathbbm{P}(x\not\sim y|x\in X)>\beta|y\in Y\right)\le\frac{p}{\beta}
\end{equation}
it follows that
\begin{equation}
\label{eq:c}
\mathbbm{P}(x\not\sim y|x\in X,y\in Y)\le p.
\end{equation}
We assume now the contrary of (\ref{eq:p}), i.e.,
\begin{equation}
\label{eq:d}
\mathbbm{P}\left(\mathbbm{P}(x\not\sim y|x\in X)>\beta|y\in Y\right)>\frac{p}{\beta}
\end{equation}
from which follows
\begin{equation}
\label{eq:e}
\mathbbm{P}(x\not\sim y|x\in X,y\in Y)>p,
\end{equation}
which is a contradiction to (\ref{eq:c}) and, therefore, concludes the proof.

\subsection*{Truncating the Hilbert space of a continuous-variable-light mode}

We show how large the Hilbert space must be to describe a continuous-variable-light mode with bounded energy, i.e., bounded photon number. Let $\rho$ be the state of interest, $N_{\rm mean}$ its mean photon number, and ${\rho}_{\rm trunc}$ the truncation of $\rho$ to the first $N$ Fock layers which is not normalized
\begin{eqnarray}
	N_{\rm mean}&=&\sum_{n=0}^\infty n\,\rho_{n,n}\ge(N+1)\sum_{n=N+1}^\infty\rho_{n,n}\nonumber\\
	&\ge&(N+1)\Tr(\rho_{\rm trunc}-\rho).
\end{eqnarray}
From this we obtain
\begin{equation}
\label{eq:traceerror}
	\Tr(\rho-\rho_{\rm trunc})\le\frac{N_{\rm mean}}{N+1}.
\end{equation}
To get from (\ref{eq:traceerror}) an error to the $1$-norm we need
\begin{lemma}[Truncation of matrices]
\label{lem:tracelem}
Let $M$ be a positive semidefinite matrix, or a trace-class operator, written as
\begin{equation}
\label{eq:Mdef}
M=\left(\begin{array}{cc}A & B \\ B^\dag & C\end{array}\right),
\end{equation}
where $A$ and $C$ are square matrices. It is true that 
\begin{equation}
\label{eq:tracelem}
\|B\|_1^2\le\|A\|_1\|C\|_1. 
\end{equation}
\end{lemma}
Inserting (\ref{eq:tracelem}) with $M=\rho$ into (\ref{eq:traceerror}) and employing the triangle inequality yields with $\|A\|_1\le 1$.
\begin{equation}
\label{eq:tracenormtrunc}
\|\rho_{\rm trunc}-\rho\|_1\le\frac{N_{\rm mean}}{N+1}+2\sqrt{\frac{N_{\rm mean}}{N+1}}\le 3\sqrt{\frac{N_{\rm mean}}{N+1}},
\end{equation}
as long as $N+1\ge N_{\rm mean}$.

\subsection*{Proof of Lemma \ref{lem:tracelem} }

We decompose the Hilbert space according to the block structure of (\ref{eq:Mdef}) as $E\oplus F$ and write $M$ as $M=\sum_k\lambda M_k$ where the $M_k$ are rank one projectors with $A_k$, $B_k$, and $C_k$ as in (\ref{eq:Mdef}) and $\lambda\ge 0$. Now, we write $\lambda M_k=|\Psi_k\rangle\langle\Psi_k|$ with $|\Psi_k\rangle=a_k|\phi_k\rangle+b_k|\psi_k\rangle$ where $|\phi_k\rangle\in E$ and $|\psi_k\rangle\in F$. From this, one obtains immediately
\begin{equation}
\label{eq:app2}
\|B_k\|_1^2=|a_k|^2|b_k|^2=\|A_k\|_1\|C_k\|_1.
\end{equation}
To conclude the proof, we write
\begin{align}
\|B\|_1\le&\sum_k\|B_k\|_1\le\sum_k\sqrt{\|A_k\|_1}\sqrt{\|C_k\|_1}\nonumber\\
\le&\sqrt{\sum_k\|A_k\|_1}\sqrt{\sum_k\|C_k\|_1}=\sqrt{\|A\|_1}\sqrt{\|C\|_1},
\end{align}
where we have used the Cauchy-Schwarz inequality.

\subsection*{Proof of equation (\ref{eqn-onb})}

It remains to consider the case where $|\alpha| \geq 2\sigma$. We start by bounding the matrix elements of the displacement operator $D(\alpha)$:
\begin{equation}
\bra{k} D(\alpha) \ket{\ell}
 = e^{-|\alpha|^2/2} \bra{k} e^{\alpha a^\dg} e^{-\alpha^* a} \ket{\ell},
\end{equation}
\begin{equation}
e^{-\alpha^* a} \ket{\ell}
 = \sum_{i=0}^\ell \tfrac{(-\alpha^*)^i}{i!} \sqrt{\ell\cdots(\ell-i+1)} \ket{\ell-i}
 = \sum_{i=0}^\ell \tfrac{(-\alpha^*)^{\ell-i}}{(\ell-i)!} \sqrt{\ell\cdots(i+1)} \ket{i},
\end{equation}
\begin{equation}
\bra{k} e^{\alpha a^\dg}
 = \sum_{j=0}^k \tfrac{\alpha^j}{j!} \sqrt{k\cdots(k-j+1)} \bra{k-j}
 = \sum_{j=0}^k \tfrac{\alpha^{k-j}}{(k-j)!} \sqrt{k\cdots(j+1)} \bra{j},
\end{equation}
\begin{equation}
\bra{k} D(\alpha) \ket{\ell}
 = e^{-|\alpha|^2/2} \sum_{j=0}^{\min(k,\ell)} \tfrac{\alpha^{k-j}}{(k-j)!} \tfrac{(-\alpha^*)^{\ell-j}}{(\ell-j)!} \sqrt{k\cdots(j+1)} \sqrt{\ell\cdots(j+1)}.
\end{equation}
Using the Cauchy-Schwarz inequality, and the binomial theorem, 
\begin{equation}
\begin{split}
\bigl| \bra{k} D(\alpha) \ket{\ell} \bigr|
 &\leq e^{-|\alpha|^2/2} 
 \Bigl[ \sum_{j=0}^{\min(k,\ell)} \Bigl( \tfrac{|\alpha|^{k-j}}{(k-j)!} \Bigr)^2 \cdot k\cdots(j+1) \Bigr]^{1/2} 
 \Bigl[ \sum_{j=0}^{\min(k,\ell)} \Bigl( \tfrac{(|\alpha|)^{\ell-j}}{(\ell-j)!} \Bigr)^2 \cdot \ell\cdots(j+1) \Bigr]^{1/2} \\
 &= e^{-|\alpha|^2/2} 
 \Bigl[ \sum_{j=0}^{\min(k,\ell)} \tbinom{k}{j} \tfrac{|\alpha|^{2(k-j)}}{(k-j)!} \Bigr]^{1/2} 
 \Bigl[ \sum_{j=0}^{\min(k,\ell)} \tbinom{\ell}{j} \tfrac{|\alpha|^{2(\ell-j)}}{(\ell-j)!} \Bigr]^{1/2} \\
 &\leq e^{-|\alpha|^2/2} 
 \Bigl[ \sum_{j=0}^k \tbinom{k}{j} |\alpha|^{2(k-j)} \Bigr]^{1/2} 
 \Bigl[ \sum_{j=0}^\ell \tbinom{\ell}{j} |\alpha|^{2(\ell-j)} \Bigr]^{1/2} \\
 &= e^{-|\alpha|^2/2} (1+|\alpha|^2)^{k/2} (1+|\alpha|^2)^{\ell/2}.
\end{split}
\end{equation}
Note that, for any fixed $k$ and $\ell$, this quantity decays exponentially as $|\alpha|$ becomes large.

We now consider the $N$-photon truncated operator $D_N(\alpha)$. We can bound it in $2$-norm as follows:
\begin{equation}
\begin{split}
\norm{D_N(\alpha)}_2
 &\leq e^{-|\alpha|^2/2} \Bigl[ \sum_{k,\ell=0}^N (1+|\alpha|^2)^k (1+|\alpha|^2)^\ell \Bigr]^{1/2} \\
 &= e^{-|\alpha|^2/2} \sum_{k=0}^N (1+|\alpha|^2)^k
 = e^{-|\alpha|^2/2} \frac{(1+|\alpha|^2)^{N+1}-1}{(1+|\alpha|^2)-1}
 \quad \text{(since $|\alpha|>0$)} \\
 &\leq e^{-|\alpha|^2/2} (1+|\alpha|^2)^{N+1} |\alpha|^{-2}
 = e^{-|\alpha|^2/2} (1+|\alpha|^2)^N (1+|\alpha|^{-2}).
\end{split}
\end{equation}
Then we can bound the scaled truncated operator $\tilde{D}_N(\alpha)$ as follows:
\begin{equation}\label{eqn-onb-step}
\begin{split}
\norm{\tilde{D}_N(\alpha)}_2
 &\leq \sqrt{2}\sigma \exp(\tfrac{|\alpha|^2}{4\sigma^2} - \tfrac{|\alpha|^2}{2}) (1+|\alpha|^2)^N (1+|\alpha|^{-2}) \\
 &= \sqrt{2}\sigma \exp[\tfrac{|\alpha|^2}{4\sigma^2} - \tfrac{|\alpha|^2}{2} + N\log(1+|\alpha|^2)] (1+|\alpha|^{-2}).
\end{split}
\end{equation}
Let 
\begin{equation}
	E: = \tfrac{|\alpha|^2}{4\sigma^2} - \tfrac{|\alpha|^2}{2} + N\log(1+|\alpha|^2)
\end{equation}	
be the quantity inside the exponential; we will upper-bound it. Note the following identity, for any $x,x_0 \in (0,\infty)$: (by approximating $\log(1+x)$ to first order at the point $x=x_0$)
\begin{equation}
\log(1+x) \leq \log(1+x_0) + \tfrac{x-x_0}{1+x_0} = \log(1+x_0) + \tfrac{1+x}{1+x_0} - 1.
\end{equation}
Set $x = |\alpha|^2$ and $x_0 = 4N$, then we have 
\begin{equation}
\log(1+|\alpha|^2) \leq \log(1+4N) + \tfrac{1+|\alpha|^2}{1+4N} - 1
 \leq \log(1+4N) + \tfrac{1+|\alpha|^2}{4N} - 1
 \leq \log(1+4N) + \tfrac{|\alpha|^2}{4N}.
\end{equation}
Then 
\begin{equation}
E \leq (\tfrac{1}{4\sigma^2} - \tfrac{1}{2} + \tfrac{1}{4}) |\alpha|^2 + n\log(1+4N).
\end{equation}
Using the fact that $\alpha \geq 2\sigma = \sqrt{8N\log(1+4N)}$, we get 
\begin{equation}
E \leq (\tfrac{1}{4\sigma^2} - \tfrac{1}{2} + \tfrac{1}{4} + \tfrac{1}{8}) |\alpha|^2
 = (-\tfrac{1}{8} + \tfrac{1}{4\sigma^2}) |\alpha|^2.
\end{equation}
Plugging this into (\ref{eqn-onb-step}), we get
\begin{equation}
\norm{\tilde{D}_N(\alpha)}_2
 \leq \sqrt{2}\sigma \exp[(-\tfrac{1}{8} + \tfrac{1}{4\sigma^2}) |\alpha|^2] (1+|\alpha|^{-2}).
\end{equation}
Using the fact that $\sigma \geq 2$ and $|\alpha| \geq 2\sigma \geq 4$, we have that
\begin{equation}
\norm{\tilde{D}_N(\alpha)}_2
 \leq \sqrt{2}\sigma \exp[-\tfrac{1}{16} |\alpha|^2] (1+|\alpha|^{-2})
 \leq \sqrt{2}\sigma \exp(-1) \tfrac{17}{16}
 < \sqrt{2}\sigma.
\end{equation}
Since the operator norm is upper-bounded by the $2$-norm, this implies (\ref{eqn-onb}).

\end{document}